# The model of a flat (Euclidean) expansive homogeneous and isotropic relativistic universe in the light of the general relativity, quantum mechanics, and observations

**Vladimír Skalský**

Faculty of Material Sciences and Technology of the Slovak University of Technology, Trnava, Slovakia; e-mail: vladimir.skalsky@stuba.sk

**Abstract** Assuming that the relativistic universe is homogeneous and isotropic, we can unambiguously determine its model and physical properties, which correspond with the Einstein general theory of relativity (and with its two special partial solutions: Einstein special theory of relativity and Newton gravitation theory), quantum mechanics, and observations, too.

## 1 Introduction

In September 28, 1905 Albert Einstein published the article *Zur Elektrodynamik bewegter Körper* (Einstein 1905) which contained the *special theory of relativity*.

Einstein in his special theory of relativity—which represents the theory of physical homogeneous and isotropic space and time—discovered the *essential connection of the four-dimensional physical homogeneous and isotropic space-time*.

This connection is expressed by

*Lorentz transformation* (*group*) (Lorentz 1904):

$$x' = \frac{x - vt}{\sqrt{1 - \frac{v^2}{c^2}}}, \tag{1a}$$

$$y' = y, \tag{1b}$$

$$z' = z, \tag{1c}$$

$$t' = \frac{t - \frac{v}{c^2}x}{\sqrt{1 - \frac{v^2}{c^2}}}, \tag{1d}$$

where $x'$, $y'$, $z'$ are the space co-ordinates and $t'$ is the time in the inertial system, which move relative to the observer at the velocity $v$; $x$, $y$, $z$ are the space co-ordinates and $t$ is the time in the observer's own inertial system.

In 1907-1915 Einstein generalised the special theory of relativity on the *phenomenon of gravity* and elaborated the *general theory of relativity*, in which he discovered the *essential connection of matter, space and time* as the *unified physical theory of matter-space-time*.

The mathematical and physical fundament of the *Einstein theory of general relativity* represents the *Einstein field equations*.

On Thursday November 25, 1915 at the meeting of the *Royal Prussian Academy of Sciences in Berlin* [1] Einstein presented the article *Die Feldgleichungen der Gravitation* (Einstein 1915), which contained the final version of

*Einstein field equations*

$$G_{im} = -\kappa\left(T_{im} - \frac{1}{2}g_{im}T\right), \tag{2}$$

where $G_{im}$ is the Einstein or conservative tensor, $\kappa$ Einstein gravitational constant [$\kappa = (8\pi G)/c^4$], $T_{im}$ energy-momentum tensor, $g_{im}$ metric or fundamental tensor, and $T$ scalar or trace of energy-momentum tensor ($T \equiv T_i^i$).

Discovery of the general relativity by Einstein has a large significance for the *cosmology*, too. Einstein was well aware of it; therefore he attempted to apply the field equations (2) to the whole *universe*.

Einstein in the spirit of tradition anticipated that the relativistic universe is *homogeneous*, *isotropic* and *static*. The field equations (2), applied to the whole homogeneous and isotropic relativistic universe do not give a static solution. Therefore, Einstein tried to modify (generalize) them so that the application to the whole homogeneous and isotropic relativistic universe would result in a static solution.

The only possible generalization of the Einstein field equations (2) which, when applied to the whole homogeneous and isotropic relativistic universe gives a static solution and does not violate the principles of general relativity, is an addition by the supplement, representing a hypothetical energy of the *physical vacuum*.

On Thursday February 8, 1917 at the meeting of the Royal Prussian Academy of Sciences in Berlin Einstein presented his article *Kosmologische Betrachtungen zur allgemeinen Relativitätstheorie* (Einstein 1917a), which contained historically first *model of the relativistic universe*, which is a solution the theoretically most possible generalised version of

*Einstein modified field equations*

$$G_{\mu\nu} - \lambda g_{\mu\nu} = -\kappa\left(T_{\mu\nu} - \frac{1}{2}g_{\mu\nu}T\right), \tag{3}$$

where $\lambda$ is the Einstein cosmological constant.

The *Einstein supplementary cosmological member* $\lambda g_{\mu\nu}$ in the Einstein modified field equations (3) can have positive, negative, or zero values, depending on the value of *Einstein adjustable cosmological constant* $\lambda$, which can obtain all hypothetically (mathematically) possible values, i.e.: $\lambda > 0$, $\lambda < 0$, or $\lambda = 0$.

The Einstein field equations (2), or (3), represent a *non-linear system of ten partial differential equations of the second order* for ten unknown functions of four variables. For their solution general method does not exist.

The Einstein theory of general relativity is logically simple, complete and unambiguously determined theory, which cannot be modified. The general theory of relativity is either valid or not, another possibility does not exist, *tertium non datur*.

---

[1] Meetings of the *Royal Prussian Academy of Sciences* in Berlin took place on Thursdays.

Einstein drew the attention to this relevant property of the general theory of relativity, in the paper *What is the theory of relativity?* which was first published in the *London Times* in November 28, 1919. Einstein wrote in it: "The chief attraction of the theory lies in its logical completeness. If a single one of the conclusions drawn from it proves wrong, it must be given up; to modify it without destroying the whole structure seems to be impossible." (Einstein 1919b).

Thirty years later in the article *On the Generalized Theory of Gravitation* Einstein wrote: "In favour of this theory are, at this point, its logical simplicity and its "rigidity". Rigidity means here that theory is either true or false, but not modifiable." (Einstein 1950).

The logical simplicity, completeness and "rigidity" of the Einstein general relativity theory are manifested in the Einstein field equations, too.

The Einstein field equations (2) contain only one adjustable parameter: the Newton gravitation constant $G$, whose value is gradually being precised on the basis of observations.

The Einstein modified field equations (3) contain two adjustable parameters: Besides the Newton gravitational constant $G$, contain even the Einstein cosmological constant $\lambda$, which can be adjusted, based on observation, or determined on the basis of any physical principle.

The Einstein theory of general relativity at present time is the most verified physical theory. By many years of observations of the binary pulsar PSR 1913+16, which was discovered in July 2, 1974 by Russell A. Hulse and Joseph H. Taylor, Jr.[2], the general relativity is verified with the uncertainty $10^{-14}$ (Hawking and Penrose 1996, p. 61; Penrose 1997, p. 26).

According to Roger Penrose "... this accuracy has apparently been limited merely by the accuracy of clocks on earth." (Hawking and Penrose 1996, p. 61).

All the predictions of the general theory of relativity—and its special partial solution: the special theory of relativity—have been confirmed.

Just one of the predictions of the general theory of relativity: prediction of the *gravitational waves* was confirmed only indirectly.

The gravitation field on the Earth and in its near surroundings is relatively weak. The velocities of matter objects on the Earth and in its near surroundings—in comparing with boundary velocity of signal propagation $c$—are relatively small. Therefore, when determining the matter-space-time properties of matter objects on the Earth and in its near surroundings in most cases we suffice with the Newton gravitation theory and the classical mechanics.

The differences of calculating mater-space-time properties in the region of Earth which we are making using the Newton gravitation theory or the classical mechanics and using the Einstein general relativity or special relativity are relative small, prevailingly irrelevant, or even—using common measuring instruments—non-measurable. For example—according to the general theory of relativity—the relativistic mass of matter objects on the surface of Earth, as a result of the local gravitational field, is higher about approximately $7 \times 10^{-10}$ of their own (rest, Newtonian or classical-mechanical) mass.

However, if we are to achieve results with accuracy provided by current top observational technology, the Newton gravitational theory and the classical mechanics (which abstracts from relativistic effects) is not sufficient. In these cases we have to take into account general-relativistic effects caused by the local gravitational field, and with moving physical objects we have to take into account special-relativistic effects.

At present time the special-relativistic and general-relativistic effects are not only a matter of physical observations and experiments, but they are exploited in some high technologies. One of them is for example the American satellite navigation system in common commercial use, best known on the acronym *GPS* (*Global Positioning System*).

---

[2] Russell A. Hulse and Joseph H. Taylor, Jr. were "for the discovery of a new type of pulsar, a discovery that has opened up new possibilities for the study of gravitation" awarded the Nobel Prize in Physics 1993.

## 2 The equations of homogeneous and isotropic relativistic universe dynamics

The mathematical-physical fundament of the *relativistic cosmology* is represented by the *Friedmann equations of the homogeneous and isotropic relativistic universe dynamics* (Friedmann 1922, 1924), which—using the *Robertson-Walker metrics* (Robertson 1935, 1936a, b; Walker 1936)—can be expressed in the following form:

$$\dot{a}^2 = \frac{8\pi G \rho a^2}{3} - kc^2 + \frac{\Lambda a^2 c^2}{3}, \tag{4a}$$

$$2a\ddot{a} + \dot{a}^2 = -\frac{8\pi G p a^2}{c^2} - kc^2 + \Lambda a^2 c^2, \tag{4b}$$

$$p = w\varepsilon, \tag{4c}$$

where $a$ is the gauge factor, $\rho$ mass density, $k$ curvature index, $\Lambda$ cosmological constant, $p$ pressure, $w$ state equation constant, and $\varepsilon$ energy density.

The relativistic cosmology is based on the assumption of the homogeneous and isotropic distribution of matter objects in space. "The homogeneity and isotropy of the space means that we can choose such a cosmological time that in each moment the space metrics is the same in all of its points and in all directions." (Landau and Lifshitz 1988, p. 458).

There exist only three geometric spaces of constant curvature space:

a) *Spherical* (*Riemannian*) *geometric space* with constant positive space curvature.

b) *Hyperbolic* (*Lobachevskian*) *geometric space* with constant negative space curvature.

c) *Flat* (*Euclidean*) *geometric space* with constant zero space curvature.

The FRW equation (4a), (4b) and (4c) are an application of the Einstein modified field equations (3) for all three geometrical spaces with constant curvature space, i.e. they have solutions with curvature index $k = +1$, $k = -1$, and $k = 0$; with all mathematically possible values of cosmological constant $\Lambda$, i.e. with $\Lambda > 0$, $\Lambda < 0$ and $\Lambda = 0$; and with all mathematically possible values of state equation constant $w$, i.e. with $w > 0$, $w < 0$ and $w = 0$.

The logical simplicity, completeness and "rigidity" of the Einstein theory of general relativity, combined with the metrics with constant curvature of space, gives possibility on the unambiguously theoretical determination of the model and physical properties of the homogeneous and isotropic relativistic universe.

It follows from these facts:

Neither the Einstein field equations (2), nor the Einstein modified field equations (3), applied to the whole relativistic universe, do not give a static solution; therefore, the relativistic universe principally cannot be *static*.[3]

The FRW equations (4a), (4b) and (4c) with the values of the curvature index $k = +1$, $k = -1$, $k = 0$, the cosmological constant $\Lambda > 0$, $\Lambda < 0$, $\Lambda = 0$, and the state equation constant $w > 0$, $w < 0$, $w = 0$, describe an infinite number of the hypothetical homogeneous and isotropic relativistic universes in a linear approximation, in which we abstract from their relativistic properties, but do not abstract from their expansion velocity. This is the fact which allows theoretically to identify (select) unambiguously from an infinite set of mathematically possible solutions of the FRW equations of the linearized model of expansive homogeneous and isotropic relativistic universe, describing the expansive homogeneous and isotropic relativistic universe in the first (linear) approximation.

---

[3] Einstein in 1917 on the base of the modified field equations (3) constructed a *model of the spherical static homogeneous and isotropic relativistic universe* (Einstein 1917a). However, Arthur S. Eddington in the article *On the Instability of Einstein's Spherical World* (Eddington 1930) showed that not even Einstein modified field equations (3), applied to the whole homogeneous and isotropic relativistic universe, do not give static, but only *quasi-static* solution, because the Einstein model of spherical static homogeneous and isotropic relativistic universe is extremely unstable, therefore, any small fluctuation converted it into a *dynamic*.

According to the special theory of relativity, the matter objects can expand at velocity $v$ in the interval $(0, c)$, therefore, the dynamic homogeneous and isotropic relativistic universe which expands in finite distances at velocities $v \leq c$, principally cannot be *infinite*.

The *finite* dynamic homogeneous and isotropic relativistic universe is (must be) closed in space-time manner, therefore, in principle, it cannot be *contractile*.

In the *expansive* homogenous and isotropic relativistic universe the energy density of matter objects decrease, the energy density of the hypothetical physical vacuum energy, determined by the cosmological constant $\lambda$, does not change. Therefore, the law of energy-momentum conservation is valid in it only when the $\lambda = 0$.

All models of hypothetical *spherical* (Riemannian) expansive homogeneous and isotropic relativistic universes, which are the solution of the FRW equations (4a), (4b) and (4c) with $k = +1$ and $\Lambda = 0$, have the total dimensionless density of matter objects $\Omega_{tot} > 1$.

From the Schwarzschild solution of the Einstein's field equations (Schwarzschild 1916) follow unambiguously that the hypothetical expansive homogeneous and isotropic relativistic universes with the total dimensionless density of matter objects $\Omega_{tot} > 1$ would in the initial period of its expansive evolution have to expand at velocities $v > c$.

However, according to the special theory of relativity, the matter objects in principle cannot expand at the hyper-velocities.

It means that an expansive homogeneous and isotropic relativistic universe in principle cannot have total dimensionless density $\Omega_{tot} > 1$, i.e. in the first (linear) model approximation, it cannot have constant positive curvature of space, determined by the curvature index $k = +1$.

The volumes of spaces of the hypothetical *hyperbolic* (Lobachevskian) relativistic universes with the negative curvature space are determined by the divergent integral, therefore, they are an infinite (Friedmann 1924). It means although that they have total dimensionless density $\Omega_{tot} < 1$, at a finite distance from observers they would expand at velocities $v > c$. But that—according to the special theory of relativity—in principle it is not possible.

It means that an expansive homogeneous and isotropic relativistic universe in the first (linear) approximation in principle cannot have a constant negative space curvature, determined by the curvature index $k = -1$.

In the model of the expansive homogeneous and isotropic relativistic universe with the *constant zero space curvature* the *Euclid geometry* is valid.

For the Euclidean sphere is valid the known relation:

$$V = \frac{4}{3}\pi r^3, \tag{5}$$

where $V$ is the volume, and $r$ radius.

For the mass $m$ of the Euclidean homogeneous matter sphere the relation:

$$m = \frac{4}{3}\pi r^3 \rho \tag{6}$$

is valid. Therefore, using the relation (6) and the relation:

$$a := r, \tag{7}$$

the relation for the mass of a *flat (Euclidean) expansive homogeneous and isotropic relativistic universe* in the first (linear, Newtonian or classical-mechanical) approximation can be determined:

$$m = \frac{4}{3}\pi a^3 \rho_c, \tag{8}$$

where $\rho_c$ is the critical mass density.

The FRW equations (4a), (4b) and (4c) fulfil the restrictive condition, determined by the relations (8), only with $k = 0$, $\Lambda = 0$ and $w = -1/3$ (Skalský 2004).

It means that the *flat* (*Euclidean*) *expansive homogeneous and isotropic relativistic universe* (*ERU*) *model*—determined by the FRW equations (4a), (4b) and (4c) with $k = 0$, $\Lambda = 0$, and $w = -1/3$ (Skalský 1991)—*is the only one model of the expansive homogeneous and isotropic relativistic universe with the flat* (*Euclidean*) *geometry* (Skalský 2004).

## 3 The model of a flat (Euclidean) expansive non-decelerative non-accelerative homogeneous and isotropic relativistic universe

Using the FRW equations (4a) and (4b) with $k = 0$, $\Lambda = 0$, and

*total zero energy state equation* (Skalský 1991)

$$p = -\frac{1}{3}\varepsilon, \tag{9}$$

we can determine the fundamental matter-space-time parameters of the ERU model, i.e. the universe model, which describes observed expansive homogeneous and isotropic relativistic–quantum-mechanical universe in the linear approximation, in which we abstract from its relativistic and quantum-mechanical properties (Skalský 1991):

$$m = \frac{c^2}{2G}a = \frac{c^3}{2G}t, \tag{10}$$

where $t$ is the (cosmological) time (age of universe).

According to the relations (10) the fundamental parameters of ERU model, i.e. the mass (of matter objects) $m$, gauge factor (radius) $a$, and (cosmological) time $t$, grow linearly.

From the relations (10) result these increases of the fundamental parameters of ERU model:

*increase of universe mass*

$$\Delta m = \frac{c^2}{2G}\Delta a = 6.732\ 97 \times 10^{26} \text{ kg m}^{-1}, \tag{11}$$

$$\Delta m = \frac{c^3}{2G}\Delta t = 2.018\ 49 \times 10^{35} \text{ kg s}^{-1}, \tag{12}$$

*increase of gauge factor*

$$\Delta a = \frac{2G}{c^2}\Delta m = 1.485\ 22 \times 10^{-27} \text{ m kg}^{-1}, \tag{13}$$

$$\Delta a = c\Delta t = 2.997\ 924\ 58 \times 10^8 \text{ m s}^{-1}, \tag{14}$$

*increase of cosmological time*

$$\Delta t = \frac{2G}{c^3}\Delta m = 4.954\ 12 \times 10^{-36} \text{ s kg}^{-1}, \tag{15}$$

$$\Delta t = \frac{1}{c}\Delta a = 3.335\ 640\ 95 \times 10^{-9} \text{ s m}^{-1}. \tag{16}$$

In the relations (10) each from three fundamental mater-space-time parameters of the ERU model—i.e. the mass of universe $m$, the gauge factor $a$, and the cosmological time $t$—is unambiguously bounded linearly with other two fundamental parameters. Through FRW equation (4a), (4b) and (4c) with $k = 0$, $\Lambda = 0$ and $w = -1/3$, each from fundamental parameters of the ERU model $m$, $a$, and $t$, is unambiguously linearly bounded and with parameters $\rho$, $p$ and $\varepsilon$. Therefore, if in the ERU model we determine the relation of any next derived parameter with an arbitrary from mentioned parameters $m$, $a$, $t$, $\rho$, $p$ and $\varepsilon$, at the same time are unambiguously determined and its relations with all other fundamental and derived parameters of the ERU model. It makes possible a simple introduction of further derived parameters of the ERU model, and gives possibility to clarify its properties.

The parameters of the ERU model, which are determined by the relations (9) and (10), can be extended by next derived parameters: about the energy (of matter objects) $E$, determined by the Einstein relation $E = mc^2$ and by the Hubble parameter $H$, determined by the relation (45).

For better transparency, the parameters of the ERU model $a$, $t$, $H$, $m$, $E$, $\rho$, $\varepsilon$, and $p$, are presented in all possible relations and variations (Skalský 2004):

$$a = ct = \frac{c}{H} = \frac{2Gm}{c^2} = \frac{2GE}{c^4} = \sqrt{\frac{3c^2}{8\pi G \rho}} = \sqrt{\frac{3c^4}{8\pi G \varepsilon}}, \qquad a^2 = -\frac{c^4}{8\pi G p}, \qquad (17a)$$

$$t = \frac{a}{c} = \frac{1}{H} = \frac{2Gm}{c^3} = \frac{2GE}{c^5} = \sqrt{\frac{3}{8\pi G \rho}} = \sqrt{\frac{3c^2}{8\pi G \varepsilon}}, \qquad t^2 = -\frac{c^2}{8\pi G p}, \qquad (17b)$$

$$H = \frac{c}{a} = \frac{1}{t} = \frac{c^3}{2Gm} = \frac{c^5}{2GE} = \sqrt{\frac{8\pi G \rho}{3}} = \sqrt{\frac{8\pi G \varepsilon}{3c^2}}, \qquad H^2 = -\frac{c^2 p}{8\pi G}, \qquad (17c)$$

$$m = \frac{c^2 a}{2G} = \frac{c^3 t}{2G} = \frac{c^3}{2GH} = \frac{E}{c^2} = \sqrt{\frac{3c^6}{32\pi G^3 \rho}} = \sqrt{\frac{3c^8}{32\pi G^3 \varepsilon}}, \qquad m^2 = -\frac{c^8}{32\pi G^3 p}, \qquad (17d)$$

$$E = \frac{c^4 a}{2G} = \frac{c^5 t}{2G} = \frac{c^5}{2GH} = c^2 m = \sqrt{\frac{3c^8}{32\pi G^3 \rho}} = \sqrt{\frac{3c^{10}}{32\pi G^3 \varepsilon}}, \qquad E^2 = -\frac{c^{10}}{32\pi G^3 p}, \qquad (17e)$$

$$\rho = \frac{3c^2}{8\pi G a^2} = \frac{3}{8\pi G t^2} = \frac{3H^2}{8\pi G} = \frac{3c^6}{32\pi G^3 m^2} = \frac{3c^8}{32\pi G^3 E^2} = \frac{\varepsilon}{c^2} = -\frac{3p}{c^2}, \qquad (17f)$$

$$\varepsilon = \frac{3c^4}{8\pi G a^2} = \frac{3c^2}{8\pi G t^2} = \frac{3c^2 H^2}{8\pi G} = \frac{3c^8}{32\pi G^3 m^2} = \frac{3c^{10}}{32\pi G^3 E^2} = c^2 \rho = -3p, \qquad (17g)$$

$$p = -\frac{c^4}{8\pi G a^2} = -\frac{c^2}{8\pi G t^2} = -\frac{c^2 H^2}{8\pi G} = -\frac{c^8}{32\pi G^3 m^2} = -\frac{c^{10}}{32\pi G^3 E^2} = -\frac{c^2 \rho}{3} = -\frac{1}{3}\varepsilon. \qquad (17h)$$

In the relations (17a)-(17h) we can see, that all (fundamental and derived) parameters of the ERU model are unambiguously linearly bounded each to other, include the relation for the pressure $p$ and the energy density (of matter objects) $\varepsilon$, representing the total zero energy state equation, which is determined by the relation (9), and presented among the relations (17h), too.

In the cosmological literature instead of the (cosmological) time (age of universe) $t$ sometimes is used and the dimensionless *conform time* $\eta$, defined by the relation:

$$\eta = \pm c \int \frac{\mathrm{d}t}{a(t)}. \tag{18}$$

Therefore, any chosen parameters of the ERU model, expressed in the dimensionless conform time $\eta$, determined by the relation (18), are shown in the Table 1.

**Table 1**

Parameters of the expansive homogenous and isotropic relativistic universe model with the total zero energy state equation $p = -\frac{1}{3}\varepsilon$ ($0 < \eta < \infty$)

| Curvature index $k$ | Gauge factor $a$ | Cosmological time $t$ | Hubble parameter $H$ | Energy density $\varepsilon$ | Dimensionless density $\Omega$ |
|---|---|---|---|---|---|
| 0 | $ct_0 e^\eta = ct$ | $t_0 e^\eta = \dfrac{a}{c}$ | $\dfrac{e^{-\eta}}{t_0} = \dfrac{1}{t}$ | $\dfrac{3c^2 e^{-2\eta}}{8\pi G t_0^2} = \dfrac{3c^2}{8\pi G}\dfrac{1}{t^2}$ | 1 |

*Note*: According to Skalský (1991)

## 4 The observed and model properties of the universe

Based on the observations at present time we reliably know that the observed *universe* at smaller cosmological distances is non-homogeneous and anisotropic, structured into a *hierarchical gravitationally bound rotating systems* (*HGRSs*) with supercritical mass density and only one resultant centre of gravity. HGRSs form (in case of neglecting the smaller systems): the *galaxies*, *clusters of galaxies* and *super clusters*.

From these facts, it results unambiguously, that in smaller cosmic distances (i.e. in the range of the largest HGRSs), the universe has the supercritical mass density. Because only under this condition can exist the HGRSs, in which the gravitational interaction of matter objects is compensated by their inertial rotational motion.

In larger cosmic distances (than are dimensions of the largest HGRSs), observed universe cannot have supercritical mass density. Because if it had supercritical mass density, HGRSs would have to exist with larger dimensions than super clusters have, i.e. they would have to exist *super-super clusters*, *super-super-super clusters* ... etc. What would we—with the present level of observational techniques—undoubtedly observe.

In larger cosmic distances (than the dimensions of the largest HGRSs), observed universe is *expansive*, *homogeneous* and *isotropic*.

At present time with relatively high accuracy we know some of the physical and model parameters of the observed universe. For example: In 2009 G. Hinshaw et al. published the article *Five-year Wilkinson Microwave Anisotropy Probe* (*WMAP*) *observations: Data processing, sky maps, and basic results* (Hinshaw et al. 2009) with the cosmological parameters derived from the WMAP measurements, and with the cosmological parameters, derived from the WMAP data combined with the distance measurements from the *Type Ia Supernovae* (*SN*) and the *Baryon Acoustic Oscillations* (*BAO*). Some of them you can see in Table 2.

**Table 2**

Selected Cosmological Parameters

| Description | Symbol | WMAP-only | WMAP+BAO+SN |
|---|---|---|---|
| Selected Parameters for Standard ΛCDM Model | | | |
| Age of universe | $t_0$ | $13.69 \pm 0.13$ Gyr | $13.72 \pm 0.12$ Gyr |
| Hubble constant | $H_0$ | $71.9^{+2.6}_{-2.7}$ km s$^{-1}$ Mpc$^{-1}$ | $70.5 \pm 1.3$ km s$^{-1}$ Mpc$^{-1}$ |
| Redshift of decoupling | $z_*$ | $1090.51 \pm 0.95$ | $1090.88 \pm 0.72$ |
| Age of decoupling | $t_*$ | $380081^{+5843}_{-5841}$ yr | $376971^{+3162}_{-3167}$ yr |
| Selected Parameter for Extended Models | | | |
| Total density | $\Omega_{tot}$ | $1.099^{+0.100}_{-0.085}$ | $1.0050^{+0.0060}_{-0.0061}$ |

*Note*: According to Hinshaw et al. (2009)

The observed *expansive homogeneous and isotropic relativistic–quantum-mechanical universe* represents a *maximum actual whole of physical reality*, which from a *macro-physical point of view* has the *relativistic properties* and from a *micro-physical point of view* has the *quantum-mechanical properties*.

The relativistic and quantum-mechanical properties are *complementary*. The *quantum-mechanical objects* (*particles*) generate the *relativistic macro-world* and vice versa, the particles can exist only in the relativistic macro-world.

The observed universe from the relativistic point of view represents the relativistic *matter-space-time* (or the *matter-spacetime*), in which the matter objects determine the properties (geometry) of the space-time, and the space-time has influence on the relativistic properties and movement of matter objects. Therefore, to complete observed physical properties (and with it unambiguously bounded model properties), of observed expansive homogeneous and isotropic relativistic–quantum-mechanical universe we need to know:

a) *total mass* (*energy*) *of the universe*,

   or:

b) *spacetime properties of the universe*.

Using the total energy of the universe, or using the spacetime properties of the universe, we can determine the model and the physical matter-spacetime properties of the observed universe.

The FRW equations (4a), (4b) and (4c) describe the models of homogeneous and isotropic relativistic universe in the first (linear) approximation.

In the observed expansive homogenous and isotropic relativistic–quantum-mechanical universe in the first (linear) approximation the Newtonian relations are valid.

In the ERU model the *Euclid geometry* is valid and the same geometry is valid in the *Newton theory of general gravitation*. The ERU model and the Newton gravitation theory describe the physical macro-world in the linear approximation (which abstracts from the relativistic and quantum-mechanical properties). It means that the *ERU model is a special partial solution of the Newton gravitational theory*.

We can convince ourselves about it:
In the Newton gravitational theory

*escape velocity*

$$v_{esc} = \sqrt{\frac{2Gm}{r}}. \tag{19}$$

If in the relation (19) for $v_{esc}$ we put the velocity $c$, we receive the relation for
*Schwarzschild critical (gravitational) radius*

$$r_c = \frac{2Gm}{c^2}. \tag{20}$$

If in the relation (20) instead $r_c$ we put the gauge factor $a$, we receive:

$$a = \frac{2Gm}{c^2},$$

given among the relations (17a).

From the relations (17a) and (20) unambiguously results: *The ERU model is also a special partial solution of the first non-trivial spherical symmetrical exterior (vacuum) solution of the Einstein field equations*, found by Karl Schwarzschild (1916).

Using the relation (6) we can rewrite the relation (20) into the form:

$$r_c = \sqrt{\frac{3c^2}{8\pi G\rho_c}}. \tag{21}$$

If in the relation (21) instead $r_c$ we put the gauge factor $a$ we get the relation:

$$a = \sqrt{\frac{3c^2}{8\pi G\rho_c}},$$

given among the relations (17a).

From the relation (21) it results:

$$\rho_c = \frac{3c^2}{8\pi G r_c^2}. \tag{22}$$

If in the relation (22) instead $r_c$ we put the gauge factor $a$, we get the relation:

$$\rho_c = \frac{3c^2}{8\pi G a^2},$$

given among the relations (17f).

According to the Einstein general relativity, for the total mass $m_{tot}$ of an arbitrary Euclidean homogeneous matter sphere with the radius $r$ is valid the relation:

$$m_{tot} = \frac{4}{3}\pi r^3 \left(\rho + \frac{3p}{c^2}\right). \tag{23}$$

In the expansive homogeneous and isotropic relativistic universe the positive energy of the matter objects is exactly compensated by their negative gravitational energy. It means that: "... the total energy of the universe is exactly zero." (Hawking 1988, p. 129). Therefore, for the total energy $E_{tot}$ and the total mass $m_{tot}$ of the expansive homogeneous and isotropic relativistic universe are valid the relations:

$$E_{tot} = m_{tot}c^2 = 0. \tag{24}$$

For the total mass of the expansive homogeneous and isotropic relativistic universe in the linear approximation $m_{tot}$—with the non-zero values of the gauge factor $a$ and the mass density $\rho$—can be valid:

$$m_{tot} = \frac{4}{3}\pi a^3 \left(\rho + \frac{3p}{c^2}\right) = 0 \qquad (25)$$

only on the condition (Skalský 2002, 2004):

$$\rho + \frac{3p}{c^2} = 0. \qquad (26)$$

For the mass density $\rho$ and the energy density $\varepsilon$ is valid the relation:

$$\varepsilon = \rho c^2, \qquad (27)$$

therefore, the relation (26) we can—using the relation (27)—rewrite into the form:

$$\varepsilon + 3p = 0. \qquad (28)$$

If in the relation (28) we express the value of pressure $p$, we receive: the total zero energy state equation

$$p = -\frac{1}{3}\varepsilon,$$

which is shown above as the relation (9) and among the relations (17h).

From the above mentioned unambiguously results: *The ERU model is only one non-formal model of the expansive homogeneous and isotropic relativistic–quantum-mechanical universe in the linear approximation with the total zero and local non-zero mass (energy).*

The ERU model, determined by the FRW equations (4a), (4b) and (4c) with $k = 0$, $\Lambda = 0$ and $w = -1/3$, is the only model of the expansive homogeneous and isotropic relativistic universe in the linear approximation with non-zero mass density $\rho$ in which the total energy $E_{tot}$ is unchanged. It means that: *The ERU model is the only non-formal model of the expansive homogeneous and isotropic relativistic universe in the linear approximation in which the law of energy conservation is valid.*

In 1973, Edward P. Tryon in the journal *Nature* published an article *Is the Universe a Vacuum Fluctuation?* in which he postulates the hypothesis according to which the observed relativistic–quantum-mechanical universe is a *vacuum fluctuation* (Tryon 1973).

The *Tryon hypothesis* is based on a combination of quantum-mechanical properties of the *physical vacuum* and mathematical-physical properties of the expansive homogeneous and isotropic relativistic universe with the total zero energy.

The ERU model is the only one non-formal model of the expansive homogeneous and isotropic relativistic–quantum-mechanical universe in the linear approximation, which has a total mass (energy) equal to zero. It means that: *The ERU model is the only model of the universe, which in the linear approximation describes the expansive homogeneous and isotropic relativistic–quantum-mechanical universe, which may be regarded as a vacuum fluctuation* (Skalský 2002).

The expansive homogeneous and isotropic relativistic–quantum-mechanical universe with the total zero mass (energy) cannot have any other acceleration than zero.

The expansion of the homogeneous and isotropic relativistic–quantum-mechanical universe in the linear model approximation conforms to the *Newton general gravity law*.

The negative acceleration (i.e. deceleration), of the matter objects $a$ on the surface of a Euclidean homogeneous matter sphere is determined by the relation:

$$\boldsymbol{a} = -\frac{Gm}{r^2}. \qquad (29)$$

If in the relations (29) we substitute the mass of the Euclidean homogeneous matter sphere $m$ by the total mass $m_{tot} = 0$ and the radius $r$ by the gauge factor $a$, we get:

$$\boldsymbol{a} = -\frac{Gm_{tot}}{a^2} = 0. \tag{30}$$

The relation (30) can be expressed using the relation (25), too.

If in the relation (30) instead $m_{tot} = 0$ we put the relation (25), we obtain:

$$\boldsymbol{a} = -\frac{4}{3}\pi G a \left(\rho + \frac{3p}{c^2}\right) = 0. \tag{31}$$

The relations (30) and (31) mathematically and physically express that what we already knew thanks to a simple, trivial consideration: The ERU with the total energy $E_{tot} = m_{tot} c^2 = 0$ throughout the whole expansive evolution expands at a constant velocity.

The acceleration $\boldsymbol{a}$ in the relation (29) can be zero only under condition that the quantity which we put instead of the mass $m$ is zero. It means that: *The ERU model is the only model of expansive homogeneous and isotropic relativistic–quantum-mechanical universe with non-zero gauge factor $a$, non-zero mass density $\rho$ and acceleration $\boldsymbol{a} = 0$.*

In the expansive homogeneous and isotropic relativistic universe with total zero-energy gravitational interaction of matter objects is compensated by their expansion, determined by the pressure $p$ in the relations (17h), i.e. the matter objects in larger distances (than are the dimensions of the largest HGRSs), are moving away from each other by a constant velocity. Therefore: *In the expansive homogeneous and isotropic relativistic universe with the total energy $E_{tot} = 0$ gravitational interaction of matter objects does not occur, it affects only their special-relativistic properties that are a result of their relative uniform rectilinear motion.*

The same conclusion we reach also by identifying the spacetime properties of the universe.

According to the observations, the universe expands in finite distances by finite velocities (Hubble 1929).

According to the *Einstein special theory of relativity* (Einstein 1905) physical objects may expand at velocities $v$ in the interval $(0, c)$. Therefore: *A homogeneous and isotropic relativistic universe in principle cannot expand at velocity $v > c$.*

*An expansive homogeneous and isotropic relativistic universe in which the physical objects in finite distances expand at velocities $v \leq c$ is finite in the space-time manner.*

From the fact that the observed finite expansive homogeneous and isotropic relativistic–quantum-mechanical universe from the relativistic point of view represents mater-space-time, unambiguously result: *The finite expansive homogeneous and isotropic relativistic universe is closed in space-time* (Einstein 1919a).

A backward extrapolation of evolution of the expansive homogeneous and isotropic relativistic universe from the relativistic point of view leads to a *geometrical point* (*"beginning" limit cosmological singularity*).

From the backward extrapolation of the universe expansion it unambiguously results: *The finite expansive homogeneous and isotropic relativistic universe can be closed in spacetime only in one possible way: by the "initial" limit cosmological singularity.*

*The finite expansive homogeneous and isotropic relativistic universe can be limit-singularly closed in space-time only if during the whole expansive evolution in the maximum (limit) distance from each observer expands at the maximum (limit) velocity of signal propagation $c$, i.e. only on the assumption if the gravitational properties of matter objects in it are exactly compensated by their expansion and due to their relative movements only their special-relativistic properties are manifested.*

These properties result also from the fact that observed expansive relativistic universe in larger distances (than the dimensions of the largest HGRSs) is homogeneous and isotropic.

*The expansive relativistic universe can be homogeneous and isotropic only on the assumption that during the whole expansive evolution it expands at the maximum possible velocity of signal propagation c. Therefore, the maximum (boundary, limit) velocity of signal propagation c is the only velocity, which is not dependent on the velocity of its source, and therefore, nor on the velocity and location of the observer.*

By the fact that in the larger distances of the expansive homogeneous and isotropic relativistic universe with the total zero energy are manifested only special-relativistic properties of matter objects all its other physical and model properties are given.

The observers in the expansive homogeneous and isotropic relativistic universe—due to the *Lorentz time dilation*, determined by the relation (1d)—are contemporaries of all cosmological times, including a limit "beginning" of the expansive evolution of the relativistic universe. It means that: *The expansion velocity of the relativistic universe is determined by the velocity at which evolution of relativistic universe expansion "began", because as a result of the Lorentz time dilation is identical with it.* Therefore: *The relativistic–quantum-mechanical universe throughout the whole expansive evolution may expand at only one possible velocity: boundary (maximum, limit) velocity of signal propagation c.*

A hypothetical universe which would expand at a velocity $v < c$, would be non-homogeneous and anisotropic, would have only one privileged centre and would not be closed in the space-time manner. Therefore, an assumption of an expansive homogeneous and isotropic relativistic universe, which expands at velocity $v < c$, represents *contradictio in adjecto*.

The observed expansive homogeneous and isotropic relativistic universe in which the gravitational interaction of material objects is compensated by their expansion is *pseudo-flat* (*pseudo-Euclidean*), i.e. it has the *Minkowski pseudo-Euclidean geometry*, which differs from *Euclidean geometry* in such a way that it is influenced by the special-relativistic effects equally straightforward expanding inertial matter objects.

The pseudo-flat (pseudo-Euclidean) expansive homogeneous and isotropic special-relativistic universe in the linear approximation is a *flat* (*Euclidean*).

In the model of the expansive homogeneous and isotropic relativistic–quantum-mechanical universe (in which we abstract from the relativistic and quantum-mechanical effects, i.e. in the model universe which describes the observed universe in linear, i.e. non-relativistic approximation), the Euclidean geometry is valid (i.e. *de facto* linearized Minkowski pseudo-Euclidean geometry, in which we abstract from the special-relativistic effects), and

*Galilean transformation:*

$$x' = x - vt, \qquad y' = y, \qquad z' = z, \qquad t' = t, \tag{32}$$

i.e. *de facto* linearized Lorentz transformation, determined by the relations (1).

For the gauge factor $a$ and the cosmological time $t$ of the homogeneous and isotropic relativistic universe with the total zero and local non-zero mass (energy), which expands at a constant maximum possible (limit) velocity $c$, is valid the relation (Skalský 1992, 1989):

$$a = ct, \tag{33}$$

which is shown among the relations (17a), too.

From the relations (7), (19) and (33) it results that matter objects in the ERU model in any distance $r \leq a$ expand at an escape velocity (Skalský 2004)

$$v_{esc} = \frac{r}{a}c. \tag{34}$$

From the relation (34) it results that the model of the ERU in the distance of gauge factor $a$ expands at the escape velocity $v_{esc} = c$, in the distance $r = a/2$ expands at escape velocity $v_{esc} = c/2$, in the distance $r = a/3$ expands at escape velocity $v_{esc} = c/3$ ... etc.

That is indeed the case, we can be persuaded by a simple calculation:

From the relations (17a), (17d) and (17f)—or from the relations (8), (36), (37) and (38)—it results that the expansive non-decelerative non-accelerative homogeneous and isotropic relativistic universe with the total energy $E_{tot} = 0$ at certain cosmological time $t$, for example at

*cosmological time*

$$t_x = 15 \text{ Gyr}, \tag{35}$$

in the first (linear) approximation will have:

*gauge factor*

$$a_x = ct_x \equiv \sqrt[3]{\frac{3m_x}{4\pi\rho_x}} = 1.419 \times 10^{26} \text{ m}, \tag{36}$$

*mass*

$$m_x = \frac{c^3 t_x}{2G} \equiv \frac{4}{3}\pi a_x^3 \rho_x = 9.554 \times 10^{52} \text{ kg}, \tag{37}$$

*mass density*

$$\rho_x = \frac{3}{8\pi G t_x^2} \equiv \frac{3m_x}{4\pi a_x^3} = 7.981 \times 10^{-27} \text{ kg m}^{-3}. \tag{38}$$

The expansive homogeneous and isotropic relativistic universe with the total zero energy in the cosmological time, determined by the relation (35), in the linear approximation will have mass density $\rho_x$, determined by the relations (38), mass $m_x$, determined by the relations (37), and in the distance of gauge factor $a_x$, determined by the relations (36), will—according to the relation (19)—expand at the escape velocity $v_{esc} = 2.997\,924\,58 \times 10^8 \text{ m s}^{-1} = c$.

The sphere with the radius $r = a_x/2 = 7.095 \times 10^{25}$ m, with mass density $\rho_x$, determined by the relations (38), have—according to the relation (8)—the mass $m = 1.194 \times 10^{52}$ kg, and—according to the relation (19)—in the distance $r = a_x/2$ expand at the escape velocity $v_{esc} = 1.498\,962\,29 \times 10^8$ m s$^{-1}$ = $c/2$ ... etc.

As mentioned above, the backward extrapolation of the evolution of expansive homogeneous and isotropic relativistic–quantum-mechanical universe leads to the "initial" limit cosmological singularity. Therefore, in the "initial period" its expansive evolution the relation (33) must be consistent also with the conditions arising from the *Planck quantum hypothesis* and from the *Heisenberg uncertainty principle*.

In the years 1897-1899 in the Royal Prussian Academy of Sciences in Berlin, Max K. E. L. Planck presented with five sequels of his article *Über irreversible Strahlungsvorgänge*.

On Thursday June 1, 1899 he presented the fifth and final sequel of the named article (Planck 1899). Planck in it, using four constants: the *Newton gravitational constant G*, *constant velocity of light in vacuum c*, *Planck quantum constant h* and *Boltzmann constant $k_B$*, determined the *fundamental physical units of mass*, *temperature*, *length* and *time* that are now named after him.

The *Planck mass $m_P$*, *Planck temperature $T_P$*, *Planck length $l_P$* and *Planck time $t_P$* at present time are presented with the following values:

$$m_P = \sqrt{\frac{\hbar c}{G}} = 2.176\,44 \times 10^{-8} \text{ kg}, \tag{39}$$

$$T_P = \frac{\sqrt{\frac{\hbar c^5}{G}}}{k_B} = 1.416\,785 \times 10^{32} \text{ K}, \tag{40}$$

$$l_P = \frac{\hbar}{m_P c} = \sqrt{\frac{\hbar G}{c^3}} = 1.616\,252 \times 10^{-35} \text{ m}, \qquad (41)$$

$$t_P = \frac{l_P}{c} = \sqrt{\frac{\hbar G}{c^5}} = 5.391\,24 \times 10^{-44} \text{ s}. \qquad (42)$$

The value of the Planck mass $m_P$, determined by the relation (39), from macro-physical point of view is very small (approximately two hundred thousandth of gram). The Planck temperature $T_P$, determined by the relation (40), is—according to the Planck quantum hypothesis—theoretically the maximum possible temperature, therefore, from point of view of its effect is gigantic (maximum possible).

The mass $m$ manifests itself *inertially* and *gravitationally*. The temperature $T$ manifests itself *repulsively* (by *pulling* or *negative pressure*). From comparison of the Planck mass $m_P$ with the Planck temperature $T_P$—taking into account quantization of the mass-space-time of the universe—it results that the expansive evolution of universe "began" at the maximum possible velocity.

This deductive conclusion confirmed and specifies the values of Planck length $l_P$, determined by the relation (41), and Planck time $t_P$, determined by the relation (42), from which result:

$$l_P = c t_P. \qquad (43)$$

If in the relation (43) instead Planck length $l_P$ we put a *Planckian gauge factor* $a_P$, defined by the relation: $a_P := l_P$, we obtain:

$$a_P = c t_P, \qquad (44)$$

which is the special partial solution of the relation (33).

Therefore, from the relations (39)-(44) it results unambiguously that *according to the Planck quantum hypothesis the universe its expansive evolution "begin" at only one possible velocity: at the boundary velocity of signal propagation c.*

In 1927 Werner Heisenberg in the article *Über den anschaulichen Inhalt der quantentheoretischen Kinematik und Mechanik* (Heisenberg 1927) postulated an *uncertainty principle*, according to which is not possible with unlimited precision determine simultaneously both the position and the momentum of any particle.

From the *Heisenberg uncertainty principle* (*relations*) it results that the particle cannot remain on certain place—because it would have an exact position and exact (i.e. zero) momentum—but it must permanently fluctuate.

The observations confirmed that if we minimise the space in which the particle can fluctuate (i.e. if we specify its position), then its fluctuations are accelerated, and—in result of the *trembling motions*—its uncertainty of momentum grows.

The universe at the "beginning" of it expansive evolution had minimum size parameters, therefore, the particles in it fluctuate at the maximum possible velocities. The result of these fluctuations was the maximum possible negative pressure, which compensated their mutual gravitational interaction and was one of the causes of the maximum possible velocity of the increase of matter-space-time of the universe.—It means that even *according to the Heisenberg uncertainty principle* (one of the fundamental principles of the quantum mechanics), *an expansive evolution of the universe "began" his expansive evolution at the only possible velocity: at the maximum possible* (*limit*) *velocity of signal propagation c*.

These deductive conclusions are confirmed by the observations, too:

In 1929 Edwin P. Hubble discovered the *expansion of the universe* (Hubble 1929).

Hubble on the basis of astronomical observations of the *nebulae* (*galaxies*) found: "… a roughly linear relation between velocities and distances among nebulae …" (Hubble 1929, p. 173).

At present time this relation is known as the *Hubble law* and is written in this form:
$$v = HR, \qquad (45)$$
where $v$ is the velocity of a cosmic distant object, $R$ is its distance, and $H$ Hubble "constant" (coefficient, parameter).

From the relations (33) and (45) it results the relation for the Hubble parameter $H$ and the cosmological time (age of universe) $t$ (Skalský 1991):
$$H = \frac{v}{R} = \frac{c}{a} = \frac{c}{ct} = \frac{1}{t}, \qquad (46)$$
shown among the relations (17c), too.

According to the WMAP measurements (Hinshaw et al. 2009):
(*present*) *age of universe*
$$t_0 = 13.69 \pm 0.13 \, \text{Gyr}, \qquad (47)$$
and
(*present value of*) *Hubble constant*
$$H_0 = 71.9^{+2.6}_{-2.7} \, \text{km s}^{-1} \, \text{Mpc}^{-1} = 69.2 - 74.5 \, \text{km s}^{-1} \, \text{Mpc}^{-1}. \qquad (48)$$

From the relations (46) and (47) result:
$$H_0 = \frac{1}{t_0} = 71.42^{+0.68}_{-0.67} \, \text{km s}^{-1} \, \text{Mpc}^{-1} = 70.75 - 72.10 \, \text{km s}^{-1} \, \text{Mpc}^{-1}. \qquad (49)$$

The value of $H_0$, determined by the relation (49), is in the frame of measurement uncertainty of the value of $H_0$, determined by the relation (48).

According to the WMAP+BAO+SN measurements (Hinshaw et al. 2009):
$$t_0 = 13.72 \pm 0.12 \, \text{Gyr}, \qquad (50)$$
and
$$H_0 = 70.5 \pm 1.3 \, \text{km s}^{-1} \, \text{Mpc}^{-1} = 69.2 - 71.8 \, \text{km s}^{-1} \, \text{Mpc}^{-1}. \qquad (51)$$

From the relations (46) and (50) result:
$$H_0 = \frac{1}{t_0} = 71.27^{+0.63}_{-0.62} \, \text{km s}^{-1} \, \text{Mpc}^{-1} = 70.65 - 71.90 \, \text{km s}^{-1} \, \text{Mpc}^{-1}. \qquad (52)$$

The maximum value of $H_0$, determined by the relation (52), differs from the maximum value of $H_0$, determined by the relation (51), by the value $+0.10 \, \text{km s}^{-1} \, \text{Mpc}^{-1}$.

From the comparison of the relations (48) and (49) and the relations (51) and (52) result, that according to the WMAP and the WMAP+BAO+SN observations, determined by the relations (47), (48), (50) and (51), the observed universe—in the frame measurement uncertainty—expands at the boundary velocity of signal propagation $c$.

According to the WMAP measurements (Hinshaw et al. 2009):
*age of decoupling*
$$t_* = 380081^{+5843}_{-5841} \, \text{yr}, \qquad (53)$$
and
*redshift of decoupling*
$$z_* = 1090.51 \pm 0.95 = 1089.56 - 1091.46. \qquad (54)$$

From the relations (54) and (77) result
*velocity of decoupling* $v_*$:
$$v_{z=1089.56} \leq v_* \leq v_{z=1091.46}, \qquad (55)$$

where $v_{z=1089.56} = \frac{1189320.1136}{1189322.1136} c = 0.999998318... c$, and $v_{z=1091.46} = \frac{1193467.8516}{1193469.8516} c$

$= 0.999998324... c$.

According to the WMAP+BAO+SN measurements (Hinshaw et al. 2009):

$$t_* = 376971^{+3162}_{-3167} \text{ yr,} \tag{56}$$

and

$$z_* = 1090.88 \pm 0.72 = 1090.16 - 1991.60. \tag{57}$$

From the relations (57) and (77) result:

$$v_{z=1090.16} \leq v_* \leq v_{z=1991.60}, \tag{58}$$

where $v_{z=1090.16} = \frac{1190629.1456}{1190631.1456} c = 0.999998320... c$, and $v_{z=1991.60} = \frac{3970453.76}{3970455.76} c$

$= 0.999999496... c$.

From the relation (55), or (58), taking into account the age of the universe, determined by the relation (47), or (50), and the age of decoupling, determined by the relation (53), or (56), it results that the WMAP and the WMAP+BAO+SN observations confirmed that the observed universe—in the frame measurement uncertainty—expands at the velocity $c$.

The expansive non-decelerative non-accelerative homogeneous and isotropic relativistic universe with the total zero energy which during the whole expansive evolution expands at the escape velocity $v_{esc} = c$, has the critical mass (energy) density, i.e.:

*total* (*dimensionless*) *density* (*of the universe*)

$$\Omega_{tot} = 1. \tag{59}$$

According to the WMAP measurements (Hinshaw et al. 2009):

$$\Omega_{tot} = 1.099^{+0.100}_{-0.085} = 1.014 - 1.199. \tag{60}$$

The value of $\Omega_{tot}$, determined by the relation (59), differs from the minimum value of $\Omega_{tot}$, determined by the relation (60), by the value $-0.014$.

According to the WMAP+BAO+SN measurements (Hinshaw et al. 2009):

$$\Omega_{tot} = 1.0050^{+0.0060}_{-0.0061} = 0{,}9989 - 1{,}0110. \tag{61}$$

The value of $\Omega_{tot}$, determined by the relation (59), is in the frame measurement uncertainty of the value of $\Omega_{tot}$, determined by the relation (61).

## 5 The model and physical properties of the expansive homogeneous and isotropic relativistic universe

The ERU model, determined by the FRW equations (4a), (4b) and (4c) with $k = 0$, $\Lambda = 0$ and $w = -1/3$, describes the expansive homogeneous and isotropic relativistic–quantum-mechanical universe in the linear approximation (in which we abstract from its relativistic and quantum-mechanical properties).

The ERU model is *flat* (*Euclidean*), however, the real expansive homogeneous and isotropic relativistic universe is a *pseudo-flat* (*pseudo-Euclidean*), i.e. it has the *Minkowski pseudo-*

*Euclidean geometry*, which differs from Euclidean geometry "only" in that, that it is influenced by the special-relativistic effects of the expanding inertial matter objects.

This fact makes it possible—by comparing the linearized (i.e. non-relativistic) properties of the ERU model and the non-linearized (i.e. special-relativistic) properties of the actual observed expansive homogeneous and isotropic relativistic universe—to get a certain idea about the relationship between them, and about possibilities of using the ERU model in the relativistic cosmology.

Probably you cannot imagine the evolution of the observed *four-dimensional pseudo-flat (pseudo-Euclidean) expansive non-decelerative non-accelerative homogeneous and isotropic relativistic universe*. However, you can imagine it without problems in the linear (Euclidean) model approximation, in which we abstract from its special-relativistic properties. Therefore, we start the comparison of the ERU model with the real pseudo-flat expansive homogeneous and isotropic relativistic universe by this 4-dimensional image:

The relativistic universe during its whole expansive evolution "expands" at a constant, maximum possible velocity of signal propagation $c$ in the distance of the gauge factor $a$.

The maximum velocity of signal propagation $c$ is not dependent on the velocity of its source and hence nor on the velocity and location of the observer. Therefore, all observers in the relativistic universe are in its "centre" and in Euclidean approximation (in which we abstract from its relativistic properties), it can be imagined as an expanding Euclidean homogenous matter sphere, whose surface is moving away from them at a constant velocity $c$.

If we separate the time component from spatial components and if we abstract from one spatial dimension, the evolution of 4-dimensional expansive homogeneous and isotropic relativistic universe in the Euclidean projection—which you have just imagined—can be presented in 3-dimensional linear (Euclidean) approximation in the form of a time cone, which we show in Figure 1.

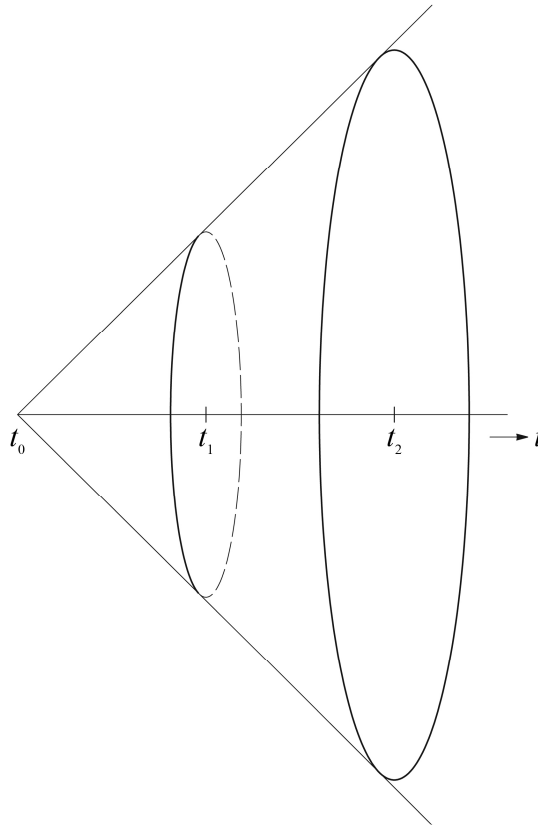

**Fig. 1** The evolution of 4-dimensional expansive homogeneous and isotropic relativistic universe in the 3-dimensional Euclidean (non-relativistic) presentation in the cosmological times $t_1$, $t_2$, ... $t_n$.

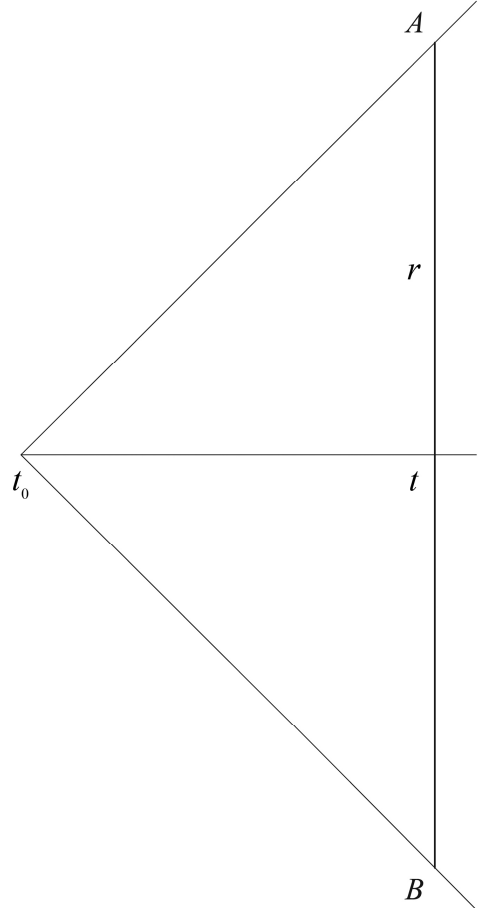

**Fig. 2** Two-dimensional Euclidean (non-relativistic) projection of the 4-dimensional expansive homogeneous and isotropic relativistic universe in an arbitrary cosmological time $t$.

The ellipses in Figure 1 represent the 2-dimensional Euclidean (non-relativistic) projection of the 3-dimensional pseudo-Euclidean space of expansive homogeneous and isotropic relativistic universe in the linear approximation in the cosmological times $t_1$, $t_2$, ... $t_n$.

If the 3-dimensional Euclidean projection of the expansive homogeneous and isotropic relativistic universe (shown in Figure 1), is reduced by another spatial dimension and the times $t_1$, $t_2$, ... $t_n$ are reduced to only one, we get 2-dimensional linearized space-time (Euclidean) projection of the 4-dimensional expansive homogeneous and isotropic relativistic universe in an arbitrary cosmological time $t$, shown in Figure 2.

In Figure 2 the abscissa connecting the point $t$ (in which is the observer), with point $A$, and point $t$ with point $B$, represent the radius of Euclidean sphere $r$, i.e. the gauge factor $a$ of the ERU model, and the abscissa connecting point $A$ with point $B$ represents the diameter of Euclidean sphere $d$, i.e. 1-dimensional model projection of the 3-dimensional space of the expansive homogeneous and isotropic relativistic universe in the linear (Euclidean) projection at any cosmological time $t$.

As mentioned earlier, the Minkowski pseudo-Euclidean geometry differs from Euclidean geometry "only" therein, that is influenced by special-relativistic effects of the inertial matter objects, expanding at constant velocities, therefore, in the place of the observer (i.e. with zero velocity of matter objects), the Minkowski pseudo-Euclidean geometry is identical with the Euclid geometry.

These facts allow us to construct 2-dimensional pseudo-Euclidean model of 4-dimensional expansive homogeneous and isotropic relativistic universe. We can do it in such a way that in the 2-dimensional Euclidean space-time model of ERU (projected in Figure 2), we take into account the special-relativistic space-time effects of expanding inertial matter objects.

From the fourth equation of the Lorentz transformation, in this article shown as the relation (1d), results dimensionless

*dilated time*

$$t' = \frac{t}{\sqrt{1-\frac{v^2}{c^2}}}. \tag{62}$$

Dimensionless proportion $v/c$ in the relation (62), representing the velocity of matter object $v$, expressed as a fraction of the velocity of light $c$—at present time prevailingly designated by letter $\beta$—is known as the dimensionless

*velocity parameter*

$$\beta = \frac{v}{c}. \tag{63}$$

The expansive homogeneous and isotropic relativistic universe throughout its whole expansive evolution expands at constant velocity $c$. For its gauge factor $a$ and the cosmological time $t$ is valid the relation (33). Therefore, for the dimensionless distance of the matter object $r$, expanding at velocity $v$ is valid the relation:

$$r = vt. \tag{64}$$

From the relations (33), (63) and (64) it results the dimensionless proportion $r/a$, representing a distance of expanding matter object $r$, expressed as a fraction of the gauge factor $a$, which represents the dimensionless

*distance parameter*

$$R = \frac{r}{a}, \tag{65}$$

expressing a linearized (non-relativistic) distance of the expanding matter object.

From the relations (33), (63), (64) and (65) it results:

$$R = \frac{r}{a} = \frac{vt}{ct} = \frac{v}{c} = \beta. \tag{66}$$

The dimensionless inverted value of the root, presented in the relation (62)—at present time mainly designed by the letter $\gamma$—is known as the dimensionless

*Lorentz factor*

$$\gamma = \frac{1}{\sqrt{1-\beta^2}}. \tag{67}$$

Using the relation (67) we can rewrite the relation (62) into the form:

$$t' = \gamma t. \tag{68}$$

As a result of the time dilation, determined by the relation (62), or by the relation (68), looking into the distance, in certain sense, we look into "the past". Strictly speaking, we observe the events, which—from observers' point of view who are located on the observed place—are already in the past.

The object expanding at an arbitrary velocity, which is determined by the dimensionless velocity parameter $\beta$, is observed in the corresponding dimensionless

*proportional time*

$$t_p = \frac{t}{t'} \equiv \frac{1}{\gamma} \equiv \sqrt{1-\beta^2}. \tag{69}$$

The dimensionless proportional time $t_p$, determined by the relations (69), is shown in Figure 3.

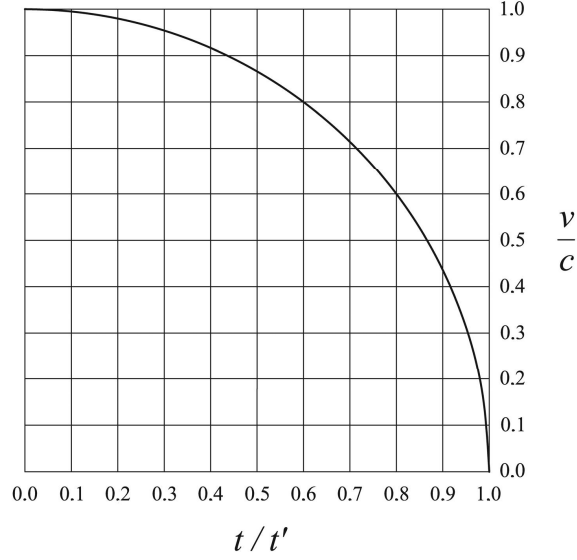

**Fig. 3** The proportional time $t_p = t/t'$ at the velocities, determined by the velocity parameter $\beta = v/c$.

In the expansive relativistic universe the distances between expanding inertial matter objects, increase proportionally to time. Therefore, if we look into the distance—due to the Lorentz time dilatation—we observe the regions of the universe in which the distance between the matter objects expanding at the same velocities are smaller. Therefore, if we project the real (non-linearized) properties of the expansive homogeneous and isotropic relativistic universe we must take into account also this fact.

The distances of the inertial matter objects, which expand at the velocities, expressed by the dimensionless velocity parameter $\beta$, determined the dimensionless

*spacetime parameter*

$$\delta = \beta t_p \equiv \frac{\beta}{\gamma} = \frac{R}{\gamma} \equiv R t_p. \tag{70}$$

In Figure 4 into 2-dimensional linearized (non-relativistic) projection of the evolution of the expansive relativistic universe, which is shown in Figure 2, we projected the proportion time $t_p$, determined by the relations (69), and shown in Figure 3. The result is a 2-dimensional projection of evolution of the 4-dimensional pseudo-Euclidean expansive homogeneous and isotropic relativistic universe, projected into 2-dimensional linearized model of ERU.

To be able to visually compare the proportional time $t_p$, shown in Figure 3, with the proportional time $t_p$, projected in the 2-dimensional linearized ERU model and shown in Figure 4, we have inserted spatial-temporal grids into Figures 3 and 4.

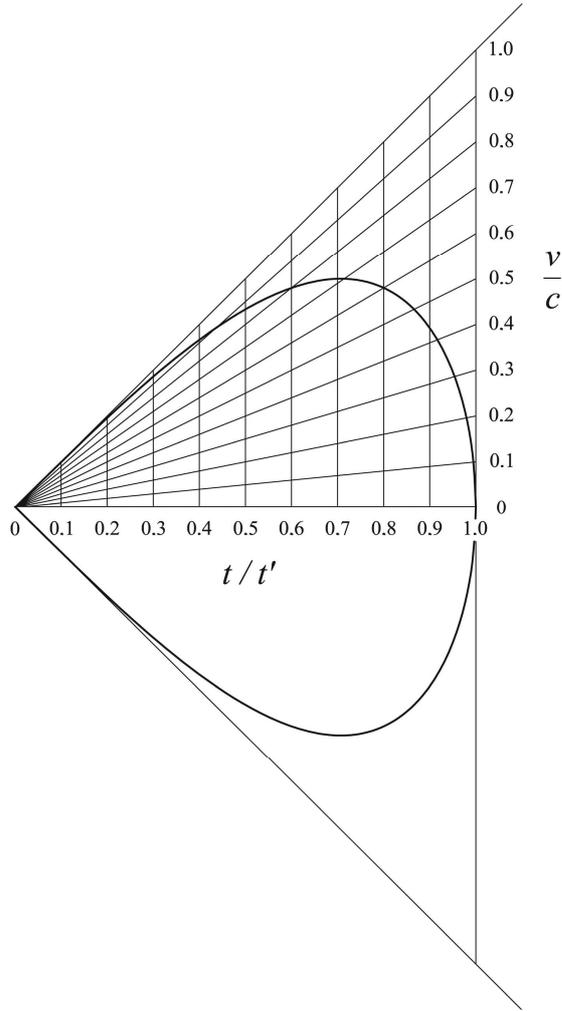

**Fig. 4** Two-dimensional projection of the 4-dimensional pseudo-Euclidean expansive homogeneous and isotropic relativistic universe.

In Figure 4 we can see that the dimensionless proportional time $t_p$, projected into the 2-dimensional linearized ERU model, expanding at velocities, expressed by dimensionless velocity parameter $\beta$, represents the dimensionless space-time parameter $\delta$, determined by the relations (70). (Compare with the values of $\delta$ in Table 3 on pages 29 and 30.)

In order to accentuate the coincidences and the differences between linearized (non-relativistic) model of evolution of expansive relativistic universe (in Figures 3 and 4 projected in 2-dimensional Euclidean projection) and the expansive relativistic universe (in Figure 4 projected in 2-dimensional model pseudo-Euclidean projection) the Figure 4 was modified (simplified and supplemented), into the form of Figure 5.

In Figures 4 and 5 we can compare the properties of the model of expansive relativistic universe, projected in the 2-dimensional linearized (flat, Euclidean, i.e. non-relativistic), model projection, with the model of the expansive relativistic universe, projected in the 2-dimensional pseudo-flat (pseudo-Euclidean) special-relativistic model projection.

In Figure 5 we can see that to the gauge factor $a$ (which connects the point $t$ with the point $A$ in the 2-dimensional linearized (i.e. Euclidean, non-relativistic) projection of the expansive homogeneous and isotropic relativistic universe, corresponds to the special-relativistic gauge factor $a'$ (which connects the point $t$—in which is the observer—with the point $t_0 \equiv A' \equiv B'$), in the 2-dimensional pseudo-Euclidean projection.

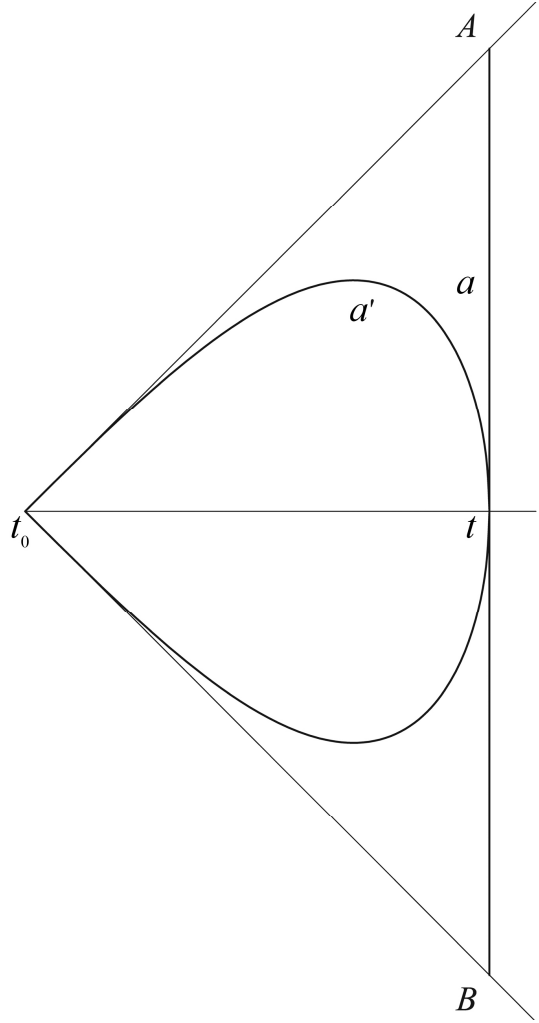

**Fig. 5** Two-dimensional projection of the 4-dimensional expansive homogeneous and isotropic relativistic universe at an arbitrary cosmological time *t*.

In Figures 4 and 5, we see that properties of the linearized ERU model and the properties of the model of expansive homogeneous and isotropic (special-)relativistic universe coincide only in the observer coordinate system (i.e. at point *t*), with the value of dimensionless velocity parameter $\beta = 0$. Because the relation (62) (or the relation (68)) gives the value $t' = t$ only at the velocity $v = 0$. From this fact it results the relation of the linearized ERU model and the expansive homogeneous and isotropic relativistic universe:

*The ERU model* (determined by the FRW equations (4a), (4b) and (4c) with $k = 0$, $\Lambda = 0$ and $w = -1/3$), *is the extrapolation of the local* (*idealised*) *properties of the expansive non-decelerative non-accelerative homogeneous and isotropic relativistic universe in the observer place to the whole universe.*

With an arbitrary small velocity *v* the linearized (Euclidean) parameters of the ERU model and special-relativistic (pseudo-Euclidean) parameters of the expansive homogeneous and isotropic relativistic universe are different.

In Figures 4 and 5 we can see that at low velocities, the differences between the dimensionless linearized cosmological time *t* and dimensionless proportional time $t_p$ are small. With bigger velocities differences nonlinearly increase and in the value of $\beta = 1$ the difference exceeds all limits.

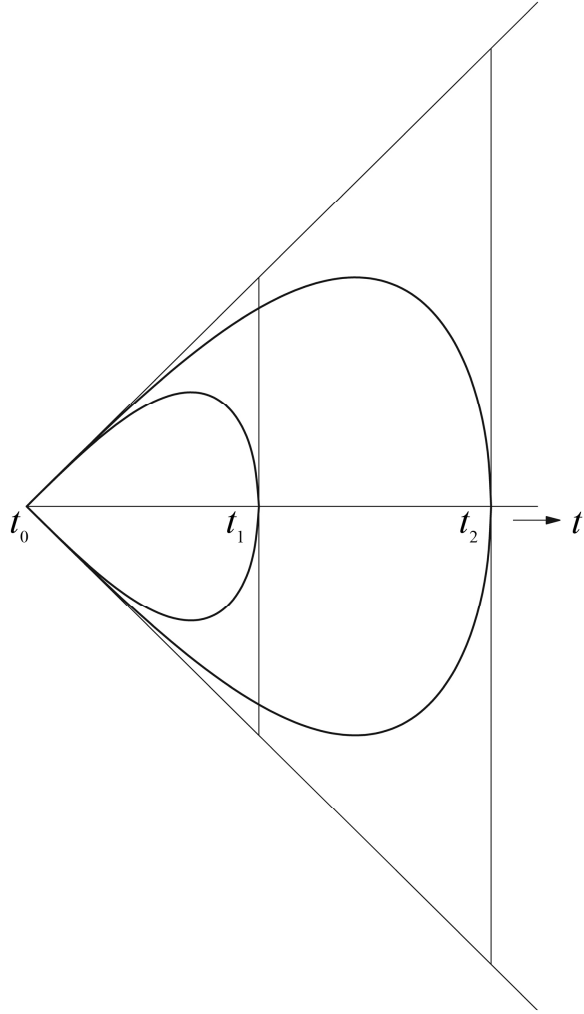

**Fig. 6**  The evolution of the 4-dimensional expansive pseudo-Euclidean relativistic universe in the 2-dimensional projection in the cosmological times $t_1, t_2 \ldots t_n$.

In order to make a conception about whole evolution of the 4-dimensional expansive homogeneous and isotropic relativistic universe, in Figure 6, we projected its evolution in the 2-dimensional projection in the cosmological times $t_1, t_2 \ldots t_n$.

The time-cone in Figure 6 represents the *horizon of (all) events*. Curves for observers in the points $t_1, t_2 \ldots t_n$ connect relatively simultaneous events, therefore, they represent the *optical horizons* (*horizons of visibility*, *horizons of particles*). The points inside the optical horizons (curves) represent the *past events*. The points between the horizon of events (time-cone) and the optical horizons (curves) are the *future events*.

In Figure 6 we can see that the pseudo-Euclidean expansive homogeneous and isotropic relativistic universe is closed in the space-time and during the whole expansive evolution in the largest (limit) distance from each observer $R'_{max} = a'$, i.e. at the point $t_0$, it "expands" at the maximum velocity of signal propagation $c$.[4]

---

[4] The figures, corresponding to the figures that are projected in this article as Figures 3, 4, 5 and 6, in 1991-2006 were published in several articles and books. Unfortunately, all these images were displayed incorrectly (they were deformed curve showing the proportional time $t_p$ and the space-time parameter $\delta$).

According to the *Einstein special theory of relativity*, the expanding objects with its own (rest) mass at a velocity $c$ would have an infinite special-relativistic mass. Therefore, the material objects with non-zero rest mass in principle cannot expand at the velocity $c$.

According to the *Planck quantum hypothesis* (Planck 1899), matter objects in the expansive universe could originate in the time $t > t_P$ (where $t_P$ is the Planck time). Therefore, although the expansive relativistic–quantum-mechanical universe in the largest (limit) distance from each observer "expands" at maximum possible (limit) velocity $c$, the matter objects with own non-zero (rest) mass in it expands at velocities $v < c$.

Einstein in his book *Über die spezielle und die allgemeine Relativitätstheorie* (*Gemeinverständlich*) in the section, dedicated to the analysis of the Lorentz transformation, contextualized that, according to the Lorentz transformation, the velocity of light $c$ is constant for all observers in all coordinate systems and has demonstrated on this example: "A light-signal is sent along the positive $x$-axis, and this light-stimulus advances in accordance with the equation

$$x = ct, \tag{71}$$

i.e. with the velocity $c$. According to the equations of the Lorentz transformation, this simple relation between $x$ and $t$ involves a relation between $x'$ and $t'$. In point of fact, if we substitute for $x$ the value $ct$ in the first and fourth equations of the Lorentz transformation, we obtain:

$$x' = \frac{(c-v)\,t}{\sqrt{1-\frac{v^2}{c^2}}},$$

$$t' = \frac{\left(1-\frac{v}{c}\right)t}{\sqrt{1-\frac{v^2}{c^2}}},$$

from which, by division, the expression

$$x' = ct' \tag{72}$$

immediately follows. … The same result is obtained for rays of light advancing in any other direction whatsoever. Of course this is not surprising, since the equations of the Lorentz transformation were derived conformably to this point of view." (Einstein 1917b, p. 23).

If in the relation (71) instead of $x$ we put the gauge factor $a$, we get the relation:

$$a = ct,$$

which we show above as the relation (33) and among the relations (17a).

If in the relation (72) instead of $x'$ we put $a'$ we receive the relation:

$$a' = ct'. \tag{73}$$

From the relations (33) and (73) result the relations:

$$\frac{a}{t} = \frac{a'}{t'} = c. \tag{74}$$

From the relations (74) it results that in the expansive homogeneous and isotropic relativistic universe the gauge factor $a$ (expressed in a linear approximation, in which we abstract from special-relativistic effects), and the special-relativistic gauge factor $a'$ (expressed in the relation to the special-relativistic dilated time $t'$), for each observer grows at the same velocity $c$.

Suppose that in the expansive relativistic universe for the special-relativistic gauge factor $a'$ and the linearized gauge factor $a$ is valid the relation:

$$a' = a. \qquad (75)$$

From the relations (74) unambiguously results that on the assumption that in the expansive homogeneous and isotropic relativistic universe is valid the relation (75), at the same time must be valid also the relation:

$$t' = t, \qquad (76)$$

and *vice versa*: if in it is valid the relation (76), at the same time must be valid in it also the relation (75).

As mentioned above, in the expansive relativistic universe in the stand-point of observer, i.e. at the velocity $v = 0$, the relation for the dilation of time (62), or (68), gives the value of dilated time $t'$, determined by the relation (76). From this fact and the relation (74) it results unambiguously that in the relativistic universe for the special-relativistic gauge factor $a'$ and the linearized gauge factor $a$ is valid the relation (75).

*The relations (75) and (76) have general validity, i.e. they are valid for all observers in all coordinate systems at any cosmological time of the expansive homogeneous and isotropic relativistic universe with the total zero mass (energy).*

The velocity of cosmic objects expansion expresses the dimensionless

*Dopplerian redshift*

$$z = \frac{1 + \frac{v}{c}}{\sqrt{1 - \frac{v^2}{c^2}}} - 1 \equiv \sqrt{\frac{c+v}{c-v}} - 1 \equiv \sqrt{\frac{1+\beta}{1-\beta}} - 1. \qquad (77)$$

From the relations (70) and (77) it results that in the expansive homogeneous and isotropic special-relativistic universe the value of dimensionless space-time parameter $\delta$ increases non-linearly from the value of $\delta = 0$ with the value of dimensionless velocity parameter $\beta = 0$ in the observer stand-point to the maximum value $\delta_{max}$, which is obtained in dimensionless

*inverse distance*

$$\delta_i \equiv \delta_{max} \equiv 0.5a' = 0.5a, \qquad (78)$$

with the value of dimensionless

*velocity parameter*

$$\beta = \sqrt{0.5} \equiv \frac{1}{\sqrt{2}} \equiv \frac{\sqrt{2}}{2} = 0.707\ 106\ 78\ ..., \qquad (79)$$

i.e. with

*radial velocity*

$$v = \sqrt{0.5}c = 2.119\ 852\ 80.\ ... \times 10^8\ \text{m s}^{-1}, \qquad (80)$$

and dimensionless

*Dopplerian red shift*

$$z = \sqrt{\frac{1+\beta}{1-\beta}} - 1 \equiv \sqrt{\frac{c+v}{c-v}} - 1 = \sqrt{2} = 1.414\ 213\ 562\ ... . \qquad (81)$$

With bigger velocities—as a consequence of the special-relativistic dilation of time, dependent on the expansion velocity of matter objects—the values of dimensionless space-time parameter $\delta$ nonlinearly decrease: from the value of inverse parameter $\delta_i = 0.5$, determined by the relations (78), with the value of the dimensionless velocity parameter $\beta$, determined by the relations (79), to the value of $\delta = 0$ with the maximum value of dimensionless velocity parameter $\beta = 1$.

In Figures 4 and 5 the special-relativistic gauge factor $a'$ (connecting the point $t$ with the point $t_0$)—as a result of linearity dilation of time $t'$ and 2-dimensional projection expansive relativistic universe (in which we separate the spatial component from the time component)—is represented by the curve. In the reality, to the special-relativistic gauge factor $a'$ correspond straight lines (abscissas) connecting the point $t$ (in which is the observer), with points $\delta_{max}$ in largest pseudo-Euclidean geometric distance with the values of velocity parameter $\beta = 0$ till $\beta = \sqrt{0.5}$ in each direction from the observer and the abscissas connecting the points $\delta_{max}$ with the point $t_0 \equiv A' \equiv B'$ with the values of the velocity parameter $\beta = \sqrt{0.5}$ till $\beta = 1$.

To be able to make a visual image, the 2-dimensional projection of the 4-dimensional large pseudo-Euclidean expansive homogeneous and isotropic relativistic universe in an arbitrary cosmological time $t$, projected in Figure 5, in Figure 7 we reduced in the 1-dimensional projection (by which we reductively eliminated geometric separation of spatial and temporal components in the 2-dimensional projection of the 4-dimensional pseudo-Euclidean space-time of the expansive homogeneous and isotropic relativistic universe).

In order to point out in the Figure 7 the relations of the 2-dimensional projection of the 4-dimensional expansive homogeneous and isotropic relativistic universe with its 1-dimensional projection, between the 2-dimensional projection of evolution of expansive homogeneous and isotropic relativistic universe (Figure 7(a)) and its 1-dimensional projection (Figure 7(c)), we projected the 2-dimensional projection (Figure 7(b)), in which we have reduced a time component by half.

In Figures 4, 5, 7(a) and 7(c) we can see that the value of the dimensionless pseudo-Euclidean space-time parameter $\delta$ with increasing velocity up to the velocity $v = \sqrt{0.5}\,c$ grows in the interval $\delta = 0$ up to $\delta_i = 0.5$. With the velocities $v > \sqrt{0.5}\,c$ the value $\delta$ decreases in the interval $\delta_i = 0.5$ till $\delta = 0$. It means that special-relativistic pseudo-Euclidean (non-linearized) distances in the expansive homogeneous and isotropic relativistic universe are determined by two relations:

a) In the interval in which the value of the dimensionless space-time parameter $\delta$ increases, i.e. in the interval special-relativistic (non-linearized) distance of expanding inertial matter objects, from the value $R' = 0$ up to the value $R' = 0.5a'$ is valid the dimensionless relation:

$$R' = \delta. \tag{82a}$$

b) In the interval in which the value of $\delta$ decreases, i.e. in the interval special-relativistic distance of the expanding matter objects, from the value $R' = 0.5a'$ up to the value $R' = a'$ is valid the dimensionless relation:

$$R' = 1 - \delta. \tag{82b}$$

To the dimensionless linearized distance of the expansive inertial matter objects $R$, determined by the relation (66), corresponds the dimensionless distance of expanding special-relativistic inertial matter objects $R'$, determined by the relation (82a) and (82b).

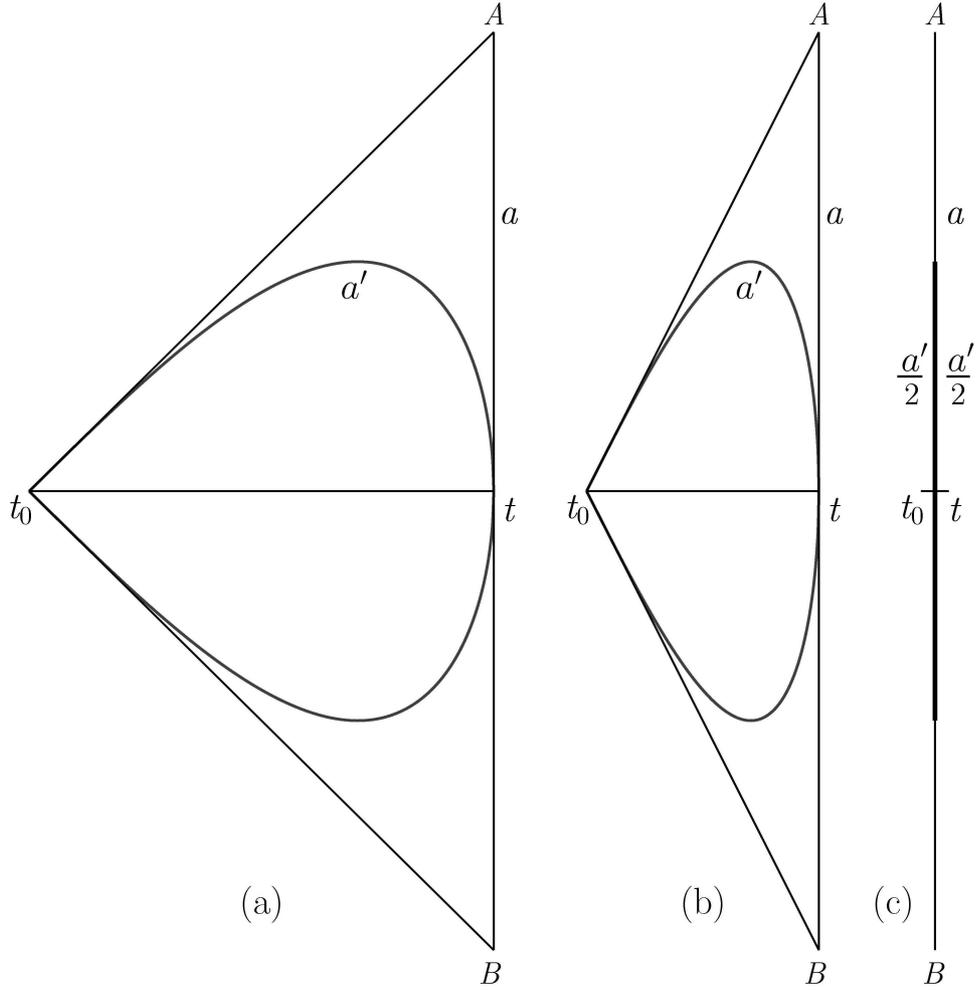

**Fig. 7** The reduction of the 2-dimensional projection of evolution of the expansive relativistic universe in an arbitrary cosmological time *t* (Fig. 7(a)), in the 1-dimensional projection (Fig. 7(c)).

For the maximum linearized distance $R_{max}$ and the (linearized) gauge factor $a$ is valid the relation:

$$R_{max} = a. \tag{83}$$

For the maximum special-relativistic (non-linearized) distance $R'_{max}$ and the (non-linearized) special-relativistic gauge factor $a'$ is valid the relation:

$$R'_{max} = a'. \tag{84}$$

Table 3 shows the values of the dimensionless Dopplerian redshift *z*, values of dimensionless Lorentz factor $\gamma$, dimensionless values of proportional time $t_p$, values of dimensionless space-time parameter $\delta$, values of dimensionless distances special-relativistic distances of expanding inertial matter objects $R'$ and values of difference distances $R - R'$ with any arbitrary velocities, expressed by means of dimensionless velocity parameter $\beta = R$.

**Table 3/1**

Selected values of some dimensionless parameters of expansive non-decelerative non-accelerative homogeneous and isotropic relativistic universe with total zero mass (energy)

| Velocity parameter $\beta = R$ | Red shift $z$ | Lorentz factor $\gamma$ | Proportional time $t_p$ | Space-time parameter $\delta$ | Distance $R'$ | Difference $R - R'$ |
|---|---|---|---|---|---|---|
| 0 | 0 | 1 | 1 | 0 | 0 | 0 |
| 0.01 | 0.01005 | 1.00005 | 0.99994 | 0.00999 | 0.00999 | $5.00013 \times 10^{-7}$ |
| 0.02 | 0.02020 | 1.00020 | 0.99979 | 0.01999 | 0.01999 | $4.0004 \times 10^{-6}$ |
| 0.03 | 0.03046 | 1.00045 | 0.99954 | 0.02998 | 0.02998 | $1.3503 \times 10^{-5}$ |
| 0.04 | 0.04083 | 1.00080 | 0.99919 | 0.03996 | 0.03996 | $3.20128 \times 10^{-5}$ |
| 0.05 | 0.05131 | 1.00125 | 0.99874 | 0.04993 | 0.04993 | $6.25391 \times 10^{-5}$ |
| 0.06 | 0.06191 | 1.00180 | 0.99819 | 0.05989 | 0.05989 | 0.00010 |
| 0.07 | 0.07263 | 1.00245 | 0.99754 | 0.06982 | 0.06982 | 0.00017 |
| 0.08 | 0.08347 | 1.00321 | 0.99679 | 0.07974 | 0.07974 | 0.00025 |
| 0.09 | 0.09444 | 1.00407 | 0.99594 | 0.08963 | 0.08963 | 0.00036 |
| 0.1 | 0.10554 | 1.00503 | 0.99498 | 0.09949 | 0.09949 | 0.00050 |
| 0.11 | 0.11677 | 1.00610 | 0.99393 | 0.10933 | 0.10933 | 0.00066 |
| 0.12 | 0.12815 | 1.00727 | 0.99277 | 0.11913 | 0.11913 | 0.00086 |
| 0.13 | 0.13967 | 1.00855 | 0.99151 | 0.12889 | 0.12889 | 0.00110 |
| 0.14 | 0.15133 | 1.00994 | 0.99015 | 0.13862 | 0.13862 | 0.00137 |
| 0.15 | 0.16315 | 1.01144 | 0.98868 | 0.14830 | 0.14830 | 0.00169 |
| 0.16 | 0.17513 | 1.01305 | 0.98711 | 0.15793 | 0.15793 | 0.00206 |
| 0.17 | 0.18728 | 1.01477 | 0.98544 | 0.16752 | 0.16752 | 0.00247 |
| 0.18 | 0.19959 | 1.01660 | 0.98366 | 0.17705 | 0.17705 | 0.00294 |
| 0.19 | 0.21207 | 1.01855 | 0.98178 | 0.18653 | 0.18653 | 0.00346 |
| 0.2 | 0.22474 | 1.02062 | 0.97979 | 0.19595 | 0.19595 | 0.00404 |
| 0.21 | 0.23759 | 1.02280 | 0.97770 | 0.20531 | 0.20531 | 0.00468 |
| 0.22 | 0.25064 | 1.02511 | 0.97549 | 0.21460 | 0.21460 | 0.00539 |
| 0.23 | 0.26388 | 1.02754 | 0.97319 | 0.22383 | 0.22383 | 0.00616 |
| 0.24 | 0.27733 | 1.03010 | 0.97077 | 0.23298 | 0.23298 | 0.00701 |
| 0.25 | 0.29099 | 1.03279 | 0.96824 | 0.24206 | 0.24206 | 0.00793 |
| 0.26 | 0.30487 | 1.03561 | 0.96560 | 0.25105 | 0.25105 | 0.00894 |
| 0.27 | 0.31898 | 1.03857 | 0.96286 | 0.25997 | 0.25997 | 0.01002 |
| 0.28 | 0.33333 | 1.04166 | 0.96 | 0.2688 | 0.2688 | 0.0112 |
| 0.29 | 0.34792 | 1.04490 | 0.95702 | 0.27753 | 0.27753 | 0.01246 |
| 0.3 | 0.36277 | 1.04828 | 0.95393 | 0.28618 | 0.28618 | 0.01381 |
| 0.31 | 0.37787 | 1.05181 | 0.95073 | 0.29472 | 0.29472 | 0.01527 |
| 0.32 | 0.39326 | 1.05550 | 0.94741 | 0.30317 | 0.30317 | 0.01682 |
| 0.33 | 0.40892 | 1.05934 | 0.94398 | 0.31151 | 0.31151 | 0.01848 |
| 0.34 | 0.42488 | 1.06334 | 0.94042 | 0.31974 | 0.31974 | 0.02025 |
| 0.35 | 0.44115 | 1.06752 | 0.93674 | 0.32786 | 0.32786 | 0.02213 |
| 0.36 | 0.45773 | 1.07186 | 0.93295 | 0.33586 | 0.33586 | 0.02413 |
| 0.37 | 0.47465 | 1.07638 | 0.92903 | 0.34374 | 0.34374 | 0.02625 |
| 0.38 | 0.49191 | 1.08109 | 0.92498 | 0.35149 | 0.35149 | 0.02850 |
| 0.39 | 0.50953 | 1.08599 | 0.92081 | 0.35911 | 0.35911 | 0.03088 |
| 0.4 | 0.52752 | 1.09108 | 0.91651 | 0.36660 | 0.36660 | 0.03339 |
| 0.41 | 0.54590 | 1.09638 | 0.91208 | 0.37395 | 0.37395 | 0.03604 |
| 0.42 | 0.56469 | 1.10189 | 0.90752 | 0.38116 | 0.38116 | 0.03883 |
| 0.43 | 0.58391 | 1.10762 | 0.90282 | 0.38821 | 0.38821 | 0.04178 |
| 0.44 | 0.60356 | 1.11358 | 0.89799 | 0.39511 | 0.39511 | 0.04488 |
| 0.45 | 0.62368 | 1.11978 | 0.89302 | 0.40186 | 0.40186 | 0.04813 |
| 0.46 | 0.64429 | 1.12622 | 0.88791 | 0.40844 | 0.40844 | 0.05155 |
| 0.47 | 0.66540 | 1.13293 | 0.88266 | 0.41485 | 0.41485 | 0.05514 |
| 0.48 | 0.68705 | 1.13990 | 0.87726 | 0.42108 | 0.42108 | 0.05891 |
| 0.49 | 0.70925 | 1.14715 | 0.87172 | 0.42714 | 0.42714 | 0.06285 |
| 0.5 | 0.73205 | 1.15470 | 0.86602 | 0.43301 | 0.43301 | 0.06698 |

**Table 3/2**

| Velocity parameter $\beta = R$ | Red shift $z$ | Lorentz factor $\gamma$ | Proportional time $t_p$ | Space-time parameter $\delta$ | Distance $R'$ | Difference $R - R'$ |
|---|---|---|---|---|---|---|
| 0.51 | 0.75545 | 1.16255 | 0.86017 | 0.43868 | 0.43868 | 0.07131 |
| 0.52 | 0.77951 | 1.17073 | 0.85416 | 0.44416 | 0.44416 | 0.07583 |
| 0.53 | 0.80425 | 1.17924 | 0.84799 | 0.44943 | 0.44943 | 0.08056 |
| 0.54 | 0.82970 | 1.18812 | 0.84166 | 0.45449 | 0.45449 | 0.08550 |
| 0.55 | 0.85592 | 1.19736 | 0.83516 | 0.45934 | 0.45934 | 0.09065 |
| 0.56 | 0.88293 | 1.20701 | 0.82849 | 0.46395 | 0.46395 | 0.09604 |
| 0.57 | 0.91080 | 1.21707 | 0.82164 | 0.46833 | 0.46833 | 0.10166 |
| 0.58 | 0.93956 | 1.22757 | 0.81461 | 0.47247 | 0.47247 | 0.10752 |
| 0.59 | 0.96927 | 1.23853 | 0.80740 | 0.47636 | 0.47636 | 0.11363 |
| 0.6 | 1 | 1.25 | 0.8 | 0.48 | 0.48 | 0.12 |
| 0.61 | 1.03179 | 1.26198 | 0.79240 | 0.48336 | 0.48336 | 0.12663 |
| 0.62 | 1.06474 | 1.27453 | 0.78460 | 0.48645 | 0.48645 | 0.13354 |
| 0.63 | 1.09890 | 1.28767 | 0.77659 | 0.48925 | 0.48925 | 0.14074 |
| 0.64 | 1.13437 | 1.30144 | 0.76837 | 0.49175 | 0.49175 | 0.14824 |
| 0.65 | 1.17124 | 1.31590 | 0.75993 | 0.49395 | 0.49395 | 0.15604 |
| 0.66 | 1.20960 | 1.33108 | 0.75126 | 0.49583 | 0.49583 | 0.16416 |
| 0.67 | 1.24957 | 1.34705 | 0.74236 | 0.49738 | 0.49738 | 0.17261 |
| 0.68 | 1.29128 | 1.36386 | 0.73321 | 0.49858 | 0.49858 | 0.18141 |
| 0.69 | 1.33486 | 1.38157 | 0.72380 | 0.49942 | 0.49942 | 0.19057 |
| 0.7 | 1.38047 | 1.40028 | 0.71414 | 0.49989 | 0.49989 | 0.20010 |
| **0.707106** | **1.414213** | **1.414213** | **0.707106** | **0.5** | **0.5** | **0.207106** |
| 0.71 | 1.42828 | 1.42004 | 0.70420 | 0.49998 | 0.50001 | 0.20998 |
| 0.72 | 1.47847 | 1.44097 | 0.69397 | 0.49966 | 0.50033 | 0.21966 |
| 0.73 | 1.53128 | 1.46317 | 0.68344 | 0.49891 | 0.50108 | 0.22891 |
| 0.74 | 1.58694 | 1.48675 | 0.67260 | 0.49772 | 0.50227 | 0.23772 |
| 0.75 | 1.64575 | 1.51185 | 0.66143 | 0.49607 | 0.50392 | 0.24607 |
| 0.76 | 1.70801 | 1.53864 | 0.64992 | 0.49394 | 0.50605 | 0.25394 |
| 0.77 | 1.77410 | 1.56729 | 0.63804 | 0.49129 | 0.50870 | 0.26129 |
| 0.78 | 1.84445 | 1.59800 | 0.62577 | 0.48810 | 0.51189 | 0.26810 |
| 0.79 | 1.91955 | 1.63103 | 0.61310 | 0.48435 | 0.51564 | 0.27435 |
| 0.8 | 2 | 1.66666 | 0.6 | 0.48 | 0.52 | 0.28 |
| 0.81 | 2.08647 | 1.70523 | 0.58642 | 0.47500 | 0.52499 | 0.28500 |
| 0.82 | 2.17979 | 1.74714 | 0.57236 | 0.46933 | 0.53066 | 0.28933 |
| 0.83 | 2.28096 | 1.79287 | 0.55776 | 0.46294 | 0.53705 | 0.29294 |
| 0.84 | 2.39116 | 1.84302 | 0.54258 | 0.45577 | 0.54422 | 0.29577 |
| 0.85 | 2.51188 | 1.89831 | 0.52678 | 0.44776 | 0.55223 | 0.29776 |
| 0.86 | 2.64495 | 1.95965 | 0.51029 | 0.43885 | 0.56114 | 0.29885 |
| **0.866025** | **2.732050** | **2** | **0.5** | **0.433012** | **0.566987** | **0.299038** |
| 0.87 | 2.79270 | 2.02818 | 0.49305 | 0.42895 | 0.57104 | 0.29895 |
| 0.88 | 2.95811 | 2.10537 | 0.47497 | 0.41797 | 0.58202 | 0.29797 |
| 0.89 | 3.14509 | 2.19317 | 0.45596 | 0.40580 | 0.59419 | 0.29580 |
| 0.9 | 3.35889 | 2.29415 | 0.43588 | 0.39230 | 0.60769 | 0.29230 |
| 0.91 | 3.60675 | 2.41191 | 0.41460 | 0.37729 | 0.62270 | 0.28729 |
| 0.92 | 3.89897 | 2.55155 | 0.39191 | 0.36056 | 0.63943 | 0.28056 |
| 0.93 | 4.25085 | 2.72064 | 0.36755 | 0.34183 | 0.65816 | 0.27183 |
| 0.94 | 4.68624 | 2.93105 | 0.34117 | 0.32070 | 0.67929 | 0.26070 |
| 0.95 | 5.24499 | 3.20256 | 0.31224 | 0.29663 | 0.70336 | 0.24663 |
| 0.96 | 6 | 3.57142 | 0.28 | 0.2688 | 0.7312 | 0.2288 |
| 0.97 | 7.10349 | 4.11345 | 0.24310 | 0.23581 | 0.76418 | 0.20581 |
| 0.98 | 8.94987 | 5.02518 | 0.19899 | 0.19501 | 0.80498 | 0.17501 |
| 0.99 | 13.10673 | 7.08881 | 0.14106 | 1.39656 | 0.86034 | 0.12965 |
| 0.999 | 43.71017 | 22.36627 | 0.04471 | 0.04466 | 0.95533 | 0.04366 |
| 1 | — | — | — | 0 | 1 | 0 |

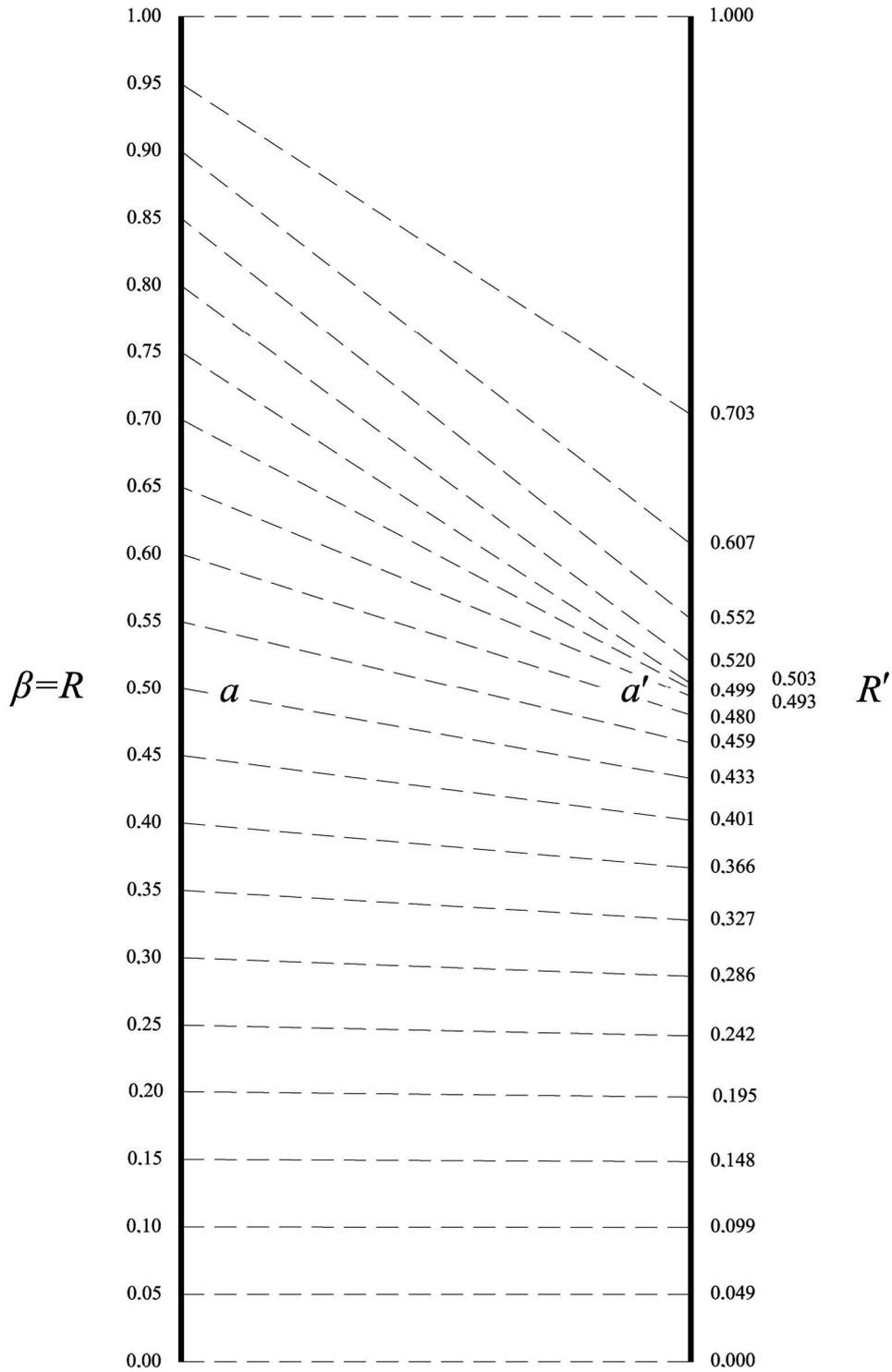

**Fig. 8** The relation of the linearized distances $R$ and the special-relativistic distances $R'$ in the expansive homogeneous and isotropic relativistic universe with some selected values of dimensionless velocity parameter $\beta$.

For illustration we show in Figure 8, the correlation of some selected linearized (non-relativistic) distances $R$ and non-linearized special-relativistic (pseudo-Euclidean) distances $R'$ with corresponding values of dimensionless velocity parameter $\beta$.

In Figure 8 and Table 3 is shown that although in the expansive homogeneous and isotropic relativistic universe for the linearized (non-relativistic) gauge factor *a* and the special-relativistic (non-linearized) gauge factor *a′* is valid the relation (75), the linearized distances *R* and special-relativistic distances *R′* have the same values only with values of dimensionless velocity parameter $\beta = 0$ and $\beta = 1$. For other velocities they have different values.

Differences between the linearized distances *R* and the special-relativistic distances *R′* with relatively low velocities are relatively small. With greater velocities up to the value of velocity parameter

$$\beta = \sqrt{0.75} \equiv \frac{\sqrt{3}}{2} = 0.866\ 025\ 403\ ...\ , \qquad (85)$$

i.e. to the linearized dimensionless distance $R = \beta = 0.866\ ...$ , which corresponds to the dimensionless special-relativistic (non-linearized) distance $R' = 0.566\ ...$ , differences between *R* and *R′* increase nonlinearly. With the value of velocity parameter $\beta = \sqrt{0.75}$ the difference $R - R'$ obtains maximum, which is approximately 0.299. With the velocities $v > \sqrt{0.75}\ c$ differences between *R* and *R′* nonlinearly decrease. Initially, only slightly, later faster. With the value of velocity parameter $\beta = 1$ the difference $R - R' = 0$.

All parameters of the pseudo-flat (pseudo-Euclidean) expansive non-decelerative non-accelerative homogeneous and isotropic relativistic universe and linearized parameters of the ERU model are mutually unambiguously bounded. Therefore, if we know any from presented special-relativistic (non-linearized) parameters, or any from linearized (i.e. non-relativistic), parameters of the expansive homogeneous and isotropic relativistic universe, with the same accuracy—with which is determined the relevant parameter—we can determine all the other (linearized and non-linearized) parameters.

## 6   The hypothetical decelerative and accelerative models of expansive relativistic universe, the ERU model, and the Type Ia supernova observations

The body with a relatively small mass (for example a satellite), in the gravitational field of a body with a relatively large mass (for example in the gravitational field of Earth) can move: a) at the velocity which is less than the escape velocity from its gravitational field; b) at the escape velocity; c) at the velocity greater than the escape velocity.

In all three cases, as a consequence of mutual gravitational interaction of matter objects, the velocity of the body is slowing down. In the first case the body moves in *elliptical* trajectory, in the second case the body moves in *parabolic* trajectory, and the in third case the body moves in *hyperbolic* trajectory.

After origin of the relativistic cosmology *analogically* there were *postulated* three hypothetical variants of so-called *decelerative model of relativistic universe:*

   a) *Model of elliptical decelerative relativistic universe* with the total dimensionless density of matter objects $\Omega_{tot} > 1$.

   b) *Model of parabolic decelerative relativistic universe* with the $\Omega_{tot} = 1$.

   c) *Model of hyperbolical decelerative relativistic universe* with the $\Omega_{tot} < 1$.

From the backward extrapolation of the expansive evolution of the relativistic universe expansion, however, it results that the expansive relativistic universe could "start" its expansive evolution only at one possible velocity: the boundary velocity of signal propagation *c*, and due to the Lorentz time dilation (1d), in maximum (limit) distance from each observer, i.e. in distance of gauge factor $a = a'$, it must expand at this velocity throughout its expansive evolution. Therefore, the expansive relativistic universe principally cannot be decelerative. (There are many other important reasons, which—because of limited space—will be not discussed in this article.)

If we ignore these facts and we assume that observed expansive homogeneous and isotropic relativistic–quantum-mechanical universe is decelerative, it would have to begin its expansive evolution at the velocity $v > c$.—However, this hypothetical decelerative assumption would be possible only with the assumption of invalidity of the theory of relativity.

This presented problem stands out best in an illustrative model projection, therefore, in Figure 9 we projected three variants of a hypothetical model of the expansive decelerative universe: (a) model of elliptical decelerative universe, (b) model of parabolic decelerative universe, (c) model of hyperbolic decelerative universe, and model (d), the model of ERU.

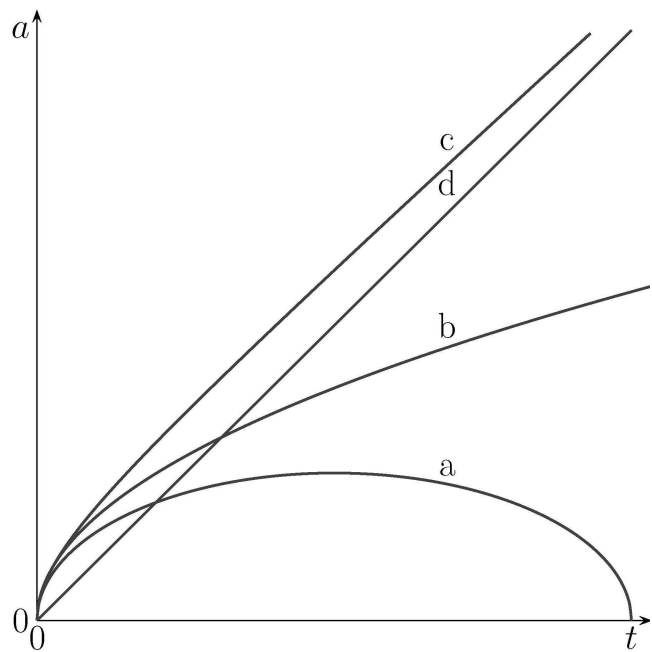

**Fig. 9** The evolution of (a) elliptical, (b) parabolic, and (c) hyperbolic model variants of a model of hypothetical expansive decelerative universe, and (d) ERU model, in two-dimensional linearized (non-relativistic) model projection.

In Figure 9 we can see that the ERU model (d) during the whole expansive evolution expands at a constant maximum possible velocity of signal propagation $c$.

All three variants of a hypothetical model of expansive decelerative relativistic universe: (a) elliptical, (b) parabolic, and (c) hyperbolic, in their initial period of expansive evolution expand at velocities $v > c$. But what is—according to the Einstein special theory of relativity—in principle not possible.

The late nineties of the last century, two international cooperating teams on the basis of observations of *Type Ia supernovae* came to a surprising conclusion: *The observed universe in the present time of its expansive evolution is not decelerative—which was at that time almost generally expected—but it is accelerative, i.e. the velocity of its expansion is not slowing but accelerating*. The results of their observations were reported in the articles: *Observational Evidence from Supernovae for an Accelerating Universe and a Cosmological Constant* (Riess et al. 1998), and *Measurements of Omega and Lambda from 42 High-Redshift Supernovae* (Perlmutter et al. 1999).

When and how was the hypothetical accelerative evolution phase of the universe to the relativistic cosmology specifically introduced, refers Robert P. Kirshner (who has been involved in the observation of the supernova since 1970), in his book *The Extravagant Universe. Exploding Stars, Dark Energy and the Accelerating Cosmos* (Kirshner 2004):

Since Hubble discovery of universe expansion in 1929, astronomers have been trying—through observations of cosmic objects—to clarify the history of its expansive evolution. But as written by Kirshner: "... toward the end of 1997 we were already beginning to see hints of something more interesting than just a low-$\Omega_m$ universe that would expand forever. Adam Riess was assembling our high-$z$ data at his office ... Adam thought he was beginning to see evidence for cosmic acceleration. Our data showed that the distant supernovae were *fainter* than they would be in a low-density universe. Faint supernovae meant larger distances. Larger distances meant cosmic acceleration. Every time he tried to use the data to determine $\Omega_m$ without $\Lambda$ the value for the mass kept coming out *negative*. That wasn't right. So he added in $\Omega_\Lambda$, and the best fit to the data points kept giving a value of the cosmological constant that was bigger than zero. As the data trickled in, Adam added more supernovae to the analysis. The statistics were beginning to make the case for the cosmological constant." (Kirshner 2004, p. 214).

Completing the model of a hypothetical expansive decelerative homogeneous and isotropic relativistic universe by a hypothetical accelerative evolution phase, a *model of an expansive decelerative-accelerative homogeneous and isotropic relativistic universe*, was introduced, now mostly known as a *standard ΛCDM model*.

During the expansive evolution of the universe the density of positive energy material objects $\varepsilon = \rho c^2$ decreases, the density of a hypothetical negative dark energy, determined by the cosmological constant $\Lambda$, however, remains unchanged. Therefore—on the assumption that at the beginning of expansive evolution of universe the absolute value of density of hypothetical negative dark energy, determined by the cosmological constant $\Lambda$, compared with positive energy density material objects $\varepsilon$, was relatively small—in the early period of expansive evolution of universe dominated the matter. Therefore—according to the standard ΛCDM model—universe in the initial period of its expansive evolution was decelerative.

The positive energy of matter objects grows more slowly than the hypothetical negative dark energy, therefore, in a certain period of expansive evolution of universe the value of positive energy and the absolute value of the hypothetical negative dark energy—according to the standard ΛCDM model—were in balance.

In the next period of expansive evolution of the universe began to dominate the hypothetical negative dark energy, therefore—according to the standard ΛCDM model—the universe in the present time of its expansive evolution is accelerative.

However, from the results of observations of WMAP and WMAP + BAO + SN (Hinshaw et al. 2009) and, from our analysis above, clearly showed that the observed expansive homogeneous and isotropic relativistic–quantum-mechanical universe in the largest (the limit) distance from each observer, i.e. in the distance of the gauge factor $a = a'$, expands at a constant maximum possible velocity of signal propagation $c$.—Of course, in such circumstances the universe cannot be nor decelerative, nor accelerative.

In Figure 8 and in Table 3 we can compare the model (linearized) and actual physical (i.e. non-linearized), parameters of the observed universe (determined on the assumption that the observed universe has the total energy $E_{tot} = 0$ and further mutually bound model and physical properties).

On the left side of Figure 8 we show the linearized dimensionless distance of expanding matter objects $R$, determined by the relation (65), linearly bound with dimensionless velocity parameters $\beta$, which are determined by the relation (63).

On the right side of Figure 8 we show the actual physical (non-linearized) dimensionless distances of matter objects $R'$, determined by the relations (82a) and (82b), which expand at the velocities, expressed by the corresponding dimensionless velocity parameters $\beta$.

In the bottom part of right side of Figure 8, between the relatively low velocities (in comparison with the boundary velocity of signal propagation $c$), and relatively small distances expanding matter objects (in comparison with the gauge factor $a = a'$), roughly linear relationships exist, as already seen in 1929 Hubble (1929).[5]

With linear growing of the velocities of expanding matter objects—as a result of expansion of the universe and special-relativistic dilation of time—their actual relativistic distances $R'$ are reduced till to the distance $a/2 = a'/2$. In bigger distances they extend. The change happens in the inverse distance $\delta_i$, determined by the relations (78), with the radial velocity, determined by the relation (80), and with the Dopplerian redshift $z$, determined by the relation (81).

Tomas Dahlen, Louis-Gregory Strolger and Adam G. Riess completed their joint paper *The Extended HST Supernova Survey: The Rate of SNe Ia at z > 1.4 Remains Low* by these Conclusions and summary: "Here we present new measurements of the Type Ia SNR to $z \sim 1.6$. Similar to our previous results based on a smaller sample, these observations show a decrease in the SNR at redshifts $z \gtrsim 1.4$, with a high significance. The results are consistent with a characteristic delay time in the order of $\tau = 2 - 3$ Gyr. Recent two-component models for the Type Ia SNR, with one dominating prompt and one less prominent delayed channel seems to fit low redshift SNR data well. These models are also consistent with a higher star formation, and they may also explain the Fe content of the inter-cluster medium in clusters of galaxies. However, these two-component models predicts rates at $z > 1.4$ that deviates from the measured rates from this investigation. Here we have discussed possible solutions to this discrepancy and found:

- It is unlikely that the difference between model predicted rates and observed rates is due to statistical fluctuations or cosmic variance.
- It is also unlikely that the low rate we measure is due to an underestimate of the host galaxy dust extinction or an overestimate detection efficiency.
- A bimodal model with a larger fraction of delayed Type Ia and that takes into account SNR hidden by dust results in a better fit to data.
- Another possible scenario that would result in a decrease in the high redshift SNR is the WD explosion efficiency deceases at high redshift." (Dahlen at al. 2008, p. 14).

The results of remote observations of Type Ia supernovae are objective, i.e.—in the frame measurements uncertainty—corresponds to the observed objective physical reality.

According to Dahlen et al. (2008), the Type Ia SNR observations show a decrease in SNR at

*redshifts*

$$z \gtrsim 1.4. \tag{86}$$

According to the observations (Hinshaw et al. 2009), the universe expands at the boundary velocity of signal propagation $c$, therefore, from comparison of the relations (81) and (86) unambiguously results that the observations of supernovae Type Ia detect the special-relativistic properties of the observed expansive homogeneous and isotropic relativistic–quantum-mechanical universe, which are a result of expansion of material objects, which compose its matter-space-time structure (and not the hypothetical acceleration of universe expansion, which contradicts to the theory of relativity and to the law of energy and momentum conservation).

---

[5] In order for Figure 8 shows that at relatively low velocities of expanding matter objects, their distances accrue not exactly linearly, we had to show them with accuracy on three decimal places.

## 7 The generally relevant laws of conservation energy, momentum, and momentum of momentum in the expansive homogeneous and isotropic relativistic universe

In the Einstein general theory of relativity (and in its two special partial solutions: Einstein special theory of relativity, and Newton theory of gravitation), and in the quantum mechanics the *generally relevant laws of conservation*: *law of energy conservation*, *law of momentum conservation*, and *law of momentum of momentum conservation* are valid.

In 1918, Emmy Noether in the article *Invariante Variationsprobleme* (Noether 1918) proved that three generally relevant laws of conservation can be expressed as the *symmetry of space and time*:

- *Law of energy conservation results from homogeneity of time.*
- *Law of momentum conservation results from homogeneity of space.*
- *Law of momentum of momentum conservation results from isotropy of space.*

From the *Noether principle of spatial and temporal symmetry of generally relevant laws of conservation* results: *Generally relevant conservation laws: the law of energy conservation, the law of momentum conservation and the law of momentum of momentum conservation are valid only in the homogeneous and isotropic universe.*

In the expansive homogeneous and isotropic relativistic–quantum-mechanical universe and in its linearized model (in which we abstract from its relativistic and quantum-mechanical properties) are valid the conservation laws of energy, momentum, and momentum of momentum, i.e. the total energy, total momentum and total momentum of momentum in it—throughout the whole expansion evolution—remain unchanged. These conditions are fulfilled by the expansive homogeneous and isotropic relativistic–quantum-mechanical universe only on the assumption that—during the whole expansive evolution—the total energy, total momentum and total momentum of momentum are equal to zero.

The matter-space-time properties of the expansive homogeneous and isotropic relativistic universe—in which are valid generally relevant laws conservation of energy, momentum and momentum of momentum—are determined by its "initial" conditions.

In 1977, Steven Weinberg in his book *The First Three Minutes: A Modern View of the Origin of the Universe* wrote: "... during the whole of the first second the universe was presumably in a state of thermal equilibrium, in which the numbers and distributions of all particles, even neutrinos, were determined by the laws of statistical mechanics, not by the details of their prior history." (Weinberg 1993, p. 146).

The properties of the *cosmic microwave background radiation*, which in the sixties of the last century were discovered by Arno A. Penzias and Robert W. Wilson (Penzias and Wilson 1965), confirmed that the observed expansive homogeneous and isotropic relativistic–quantum-mechanical universe has undergone a thermal equilibrium state.[6]

The results of observations of Penzias and Wilson specify the observation of the *COBE* (*COsmic Background Explorer*) satellite launched in November 18, 1989 by NASA (Mather et al. 1999, Smoot et al. 1992)[7] and observations of the *WMAP* (*Wilkinson Microwave Anisotropy Probe*) satellite launched in June 30, 2001 by NASA (Bennett et al. 2003, Hinshaw et al. 2009).

According to Weinberg: "When collisions or other processes bring a physical system to a state of thermal equilibrium, there are always some quantities whose values do not change. One of these "conserved quantities" is the total energy; even though collisions may transfer energy from one particle to another, they never change the total energy of the particles participating in the collision. For each such conservation law there is a quantity that must be specified before we can work out the properties of a system in thermal equilibrium—obviously, if some quantity does not change as a system approaches thermal equilibrium, but must be specified in advance. The universe has passed through a state thermal equilibrium, so to give a complete recipe for the contents of the early times, all we need is to know what were the physical quantities which were conserved as the universe expanded, and what were the values of these quantities." (Weinberg 1993, pp. 88-89).

---

[6] Arno A. Penzias and Robert W. Wilson were "for their discovery of cosmic microwave background radiation" awarded the Nobel Prize in Physics 1978.

[7] John C. Mather and George F. Smoot were "for their discovery of the blackbody form and anisotropy of the cosmic microwave background radiation" awarded the Nobel Prize in Physics 2006.

From the above analysis it results unambiguously that in the expansive homogeneous and isotropic relativistic universe is valid energy conservation law only on the assumption that its total energy is zero, i.e. only under condition that the gravitational interaction of material objects in it is exactly compensated by a negative pressure (repulsion).—This fact unambiguously determines physical and model parameters of the observed expansive homogeneous and isotropic relativistic–quantum-mechanical universe.

During the first second of the evolution of expansive homogeneous isotropic relativistic–quantum-mechanical universe due to its "initial" matter-space-time properties, "initial" temperature, Heisenberg uncertainty relations and generally relevant conservation laws of energy, momentum, and momentum of momentum there were determined its fundamental matter-space-time parameters, which in the linear approximation (in which we abstract from its relativistic and quantum-mechanical properties) are determined by the relations (17).

Einstein in the article *Die Grundlage der allgemeinen Relativitätstheorie* wrote: "... laws of conservation of momentum and energy do not apply in the strict sense for matter alone, or else that they apply only when the $g^{\mu\nu}$ are constant, i.e. when the field intensities of gravitation vanish." (Einstein 1916, p. 810).

According to the Einstein theory of gravitation: "... the absence of gravitational field implies the absence of deviation of the space-time geometry from the Euclid geometry, and also means that the curvature tensor $R^{\mu\nu}$ and its invariant $R$ are equal to zero. On the other hand, the gravitational field is absent if the mass tensor $T^{\mu\nu}$, is everywhere equal to zero. Therefore, equations $T^{\mu\nu} = 0$ and $R^{\mu\nu} = 0$ must be in any case simultaneously valid, and it is possible only if the equations conjoined $G^{\mu\nu}$ with $T^{\mu\nu}$ do not contain the member $\lambda g^{\mu\nu}$ (i.e. only when $\lambda = 0$)." (Fock 1961, p. 257).

These facts make it possible to unambiguously determine the exact solution of the Einstein modified field equations (3), applied to the whole expansive homogeneous and isotropic relativistic universe with the total zero energy (Skalský 2006):

$$G^{\mu\nu} - \lambda g^{\mu\nu} = -\kappa \left( T^{\mu\nu} - \frac{1}{2} g^{\mu\nu} T \right) = 0, \tag{87}$$

where $G^{\mu\nu} = 0$, $\lambda = 0$, $T^{\mu\nu} = 0$, and $T = 0$.

## 8 Conclusions

From the analysis given above it results unambiguously that the relativistic universe is pseudo-flat (pseudo-Euclidean) expansive non-decelerative non-accelerative homogeneous and isotropic, has the dimensionless density of matter objects $\Omega_{tot} = 1$, the energy of physical vacuum $E_{vac} = 0$, the total energy of matter objects $E_{tot} = 0$, and throughout its whole expansive evolution in the maximum (limit) distance from each observer, i.e. in the distance of gauge factor $a' = a$, expands at the constant maximum possible (limit) velocity of signal propagation $c$.

The pseudo-flat (pseudo-Euclidean) expansive non-decelerative non-accelerative homogeneous and isotropic relativistic–quantum-mechanical universe in the first (linear, Newtonian or classical-mechanical) approximation (in which we abstract from its relativistic and quantum-mechanics properties), is flat (Euclidean).

From the infinite number of theoretically possible linearized model solutions of the FRW equations (4a), (4b) and (4c) with the values of the curvature index $k = +1$, $k = -1$, $k = 0$, the cosmological constant $\Lambda > 0$, $\Lambda < 0$, $\Lambda = 0$, and the state equation constant $w > 0$, $w < 0$, $w = 0$, the above mentioned conditions, are satisfied only by one model of homogeneous and isotropic relativistic universe: the ERU model, which is their solution with the values $k = 0$, $\Lambda = 0$ and $w = -1/3$ (Skalský 1991).

If from the FRW equations (4a), (4b) and (4c) we eliminate the solutions in which the general relevant laws of conservation: law of energy conservation, law of momentum conservation, and law of momentum of momentum conservation, are invalid (i.e. the solutions that are based on hypothetical assumptions which contradict to the Einstein general theory of relativity, Einstein special theory of relativity, Newton theory of gravitation, quantum mechanics and observations), we get the final version of the linearized

*equations of the homogeneous and isotropic relativistic universe dynamics* (Skalský 1997, p. 71):

$$8\pi G a^2 \rho - 3c^2 = 0, \tag{88a}$$

$$8\pi G a^2 p + c^4 = 0, \tag{88b}$$

$$\varepsilon + 3p = 0, \tag{88c}$$

which can be described in a summary form as:

$$a^2 = \frac{3c^2}{8\pi G \rho} = -\frac{c^4}{8\pi G p}. \tag{88}$$

## 9 Afterword

The final version of the equations of the homogeneous and isotropic relativistic universe dynamics (88) describes the universe as a *givenness* in the linearized reduced form. The spacetime of the universe in them is reduced to the gauge factor $a$, and the energy of universe in them is reduced to the energy density of matter objects $\varepsilon = \rho c^2$ and the pressure $p$.

The ERU model—determined by the FRW equations (4a), (4b) and (4c) with $k = 0$, $\Lambda = 0$ and $w = -1/3$, or by the final version of the equations of dynamics of the expansive homogeneous and isotropic relativistic universe (88)—was selected from the FRW equations (4a), (4b) and (4c), on the assumptions that the observed relativistic–quantum-mechanical universe is homogeneous and isotropic.

But what was on the very "beginning" of the expansive universe evolution?
Was the universe created by the "Big Bang", or otherwise?
Or, is the universe cyclic? As, for example, the authors of cyclic an *ekpyrotic universe* Paul J. Steinhardt and Neil Turok (Steinhardt and Turok 2007) assume.
In this article we have deliberately avoided answering these and further serious questions, because the ERU model can be unambiguously determined also without their answering.